\documentclass[12pt]{article}
\usepackage[numbers,sort&compress]{natbib}
\usepackage{mciteplus}
\usepackage{grffile}
\usepackage{hyperref}
\usepackage{amsmath}
\usepackage{graphicx}
\graphicspath{{./images/}}
\usepackage{xcolor}
\RequirePackage{doi} 
\usepackage{authblk} 
\usepackage{multirow}
\usepackage{slashed}
\usepackage{float}
\usepackage{rotating}
\usepackage{caption}
\usepackage{subcaption}
\usepackage[title]{appendix}
\usepackage{dcolumn}
\newcolumntype{d}[1]{D{.}{.}{#1}}

\newcommand{\Amp}{\mathcal{M}}

\newcommand{\mycomment}[1]{}
\newcommand{\cng}[1]{\textcolor{blue}{#1}}

\newcommand{\beq}{\begin{equation}}
\newcommand{\eeq}{\end{equation}}
\newcommand{\beqa}{\begin{eqnarray}}
\newcommand{\eeqa}{\end{eqnarray}}

\newcommand{\Bbar}{\,\overline{\!B}{}}
\newcommand{\Dbar}{\,\overline{\!D}{}}
\newcommand{\Kbar}{\,\overline{\!K}{}}
\def\B0bar{\Bbar{}^0}
\def\D0bar{\Dbar{}^0}
\def\K0bar{\Kbar{}^0}

\def\Heff{\mathcal{H}_{\rm eff}}

\graphicspath{{ps}}

\newcommand{\cn}[1]{\textcolor{black}{#1}}

\def\BDlnu{$ {B} \to D^*\ell {\nu}_\ell \;$}

\title{New physics search via angular distribution of $B\to D^*\ell \nu_\ell$  decay in the light of the new lattice data}

\date{}

\author[a]{Tejhas Kapoor\thanks{\href{mailto:tejhas.kapoor@ijclab.in2p3.fr}{tejhas.kapoor@ijclab.in2p3.fr}}}
\author[b]{Zhuo-Ran Huang}
\author[a]{Emi Kou}
\affil[a]{Universit\'e Paris-Saclay, CNRS/IN2P3, IJCLab, 91405 Orsay, France}
\affil[b]{Department of Physics, College of Physics, Mechanical and Electrical Engineering, Jishou University, Jishou 416000, China}

\begin{document}
\sloppy

\maketitle

\abstract
In this article, we investigate \cn{the potential of} the angular distribution of the \BDlnu process to search for new physics signals. The Belle collaboration has analysed it to constraint $V_{cb}$ and the $B\to D^*$ form factors, under the assumption of the Standard Model. With the newly released lattice QCD data, we can perform a simultaneous fit of the form factors,  $V_{cb}$ as well as new physics parameters.  We use the Belle data and the lattice data to constrain right-handed new physics. In addition, we also generate unbinned  pseudo-dataset and perform a sensitivity study on {more general new physics} models, using the lattice data.
\vspace*{5cm}

\newpage

\section{Introduction}\label{sec:intro}
{The \BDlnu decay has been paid a great attention in recent years}. The first reason for this is the $V_{cb}$ puzzle: 
there has been a longstanding tension between the values of $|V_{cb}|$ determined using the inclusive measurements and that determined by {the} exclusive $b \rightarrow c \ell \nu$ decays. The inclusive measurement {uses the} semileptonic decays of the type $B \rightarrow X_c \ell \nu$, where $X_c$ are all possible hadronic states {including charm quark}. The current  world average is $|V_{cb}| = (42.2 \pm 0.8) \times 10^{-3}$~\cite{pdg22}. On the other hand, the exclusive measurement of $|V_{cb}|$ has the following world average: ${|V_{cb}|=}(38.46 \pm 0.40 \pm 0.55) \times 10^{-3}$, where the first uncertainty combines the statistical and systematic uncertainties from the experimental measurement and the second is theoretical (lattice QCD calculation and electroweak correction) \cite{hflav}.
\par 
The second reason { is the so-called}  $\mathcal{R}(D^{(*)})$ anomaly: to test the lepton flavour universality in the $b \rightarrow c \ell \nu$ decays,  {a} ratio is defined: 
\begin{align}
\mathcal{R}(D^{(*)}) = \frac{\mathcal{B}(B \rightarrow D^{(*)}\tau\nu)}{\mathcal{B}(B \rightarrow D^{(*)}\ell\nu)}; \quad \ell = \{ e, \mu \}
\end{align}
The combined results from BaBar~\cite{babaranomaly1,babaranomaly2}, Belle~\cite{belleanomaly1,belleanomaly2,belleanomaly3,belleanomaly4}, LHCb~\cite{lhcbanomaly1,lhcbanomaly2} and Belle II~\cite{belle2anomaly1} exceed the Standard Model (SM) predictions~\cite{hflav} by about $3.3 \sigma$. Theoretical efforts to explain such anomalies branch into model-independent and model-dependent analyses.  
For the model-dependent analysis, several classes of models have been proposed to provide a resolution to such anomalies. These classes of models can be divided into three major categories: extended higgs sector models \cite{h1,h2,h3,h4,h5,h6,h7,h8,h9,h10}, heavy charged vector boson models \cite{w1,w2,w3,w4,w5,w6} and leptoquark models \cite{tensorff,lq1,lq2,lq3,lq4,watanbeglobalfit}, with the latter being the most \cn{widely used}~\cite{anomalyrevlondon}. 
Given that the ratio $\mathcal{R}(D^{(*)})$ is measured to be higher than its SM prediction, the model-dependent approaches usually try to enhance the $B \rightarrow D^{(*)}\tau\nu$ rate (say, by introducing new heavy particles), rather than suppress the $B \rightarrow D^{(*)}\ell\nu$ rate, as the latter have stringent constraints on new physics (NP) coupling to electrons and muons~\cite{thesis}. 
\par 
In this work, {we explore the potential of the angular distribution analysis of the \BDlnu process ($\ell=e, \mu$) to search for signals beyond the SM, following the model-independent approach}. We will work with a low-energy effective Hamiltonian containing all dimension six, four-fermion operators, assuming no right-handed neutrinos. 
We expect the angular distribution to be a powerful tool to distinguish the different 
{Dirac structure that these operators have}:  (left)right-handed, (pseudo)scalar and tensor. These contributions can arise in several new NP scenarios. For example, the right-handed contribution can arise from $W_L -W_R$ mixing left-right symmetric models~\cite{lrsm1,lrsm2,lrsm3,lrsm4,lrsm5,lrsm6,lrsm7,lrsm8,lrsm9,lrsm10,lrsm11,lrsm12,lrsm13}. On the other hand, the scalar contribution is generated in models with extra charged scalars~\cite{s1,s2,s3,s4,s5,s6,s7,s8,s9,s10,s11,s12}, while the tensor operators are {often} generated by a Fierz transformation on scalar operators \cite{anomalyrevlondon}.  These NP effects can be both lepton flavour universal and lepton flavour violating, thus, having strong implication on {the $R(D^{(*)})$ anomaly \cite{marco}}. 
\par 
Belle has used the angular distribution data of \BDlnu decay to fit simultaneously $|V_{cb}|$ and the form factors~\cite{belle17,belle19,belle23}. The angular distribution is written in terms of 4 kinematical variables (one invariant mass and three angles). The analysis uses four, one-dimensional binned distributions (corresponding to each kinematical variable), in which the other three variables are integrated away. Since $|V_{cb}|$ is always multiplied with a scaling factor $\mathcal{F}(w)\eta_{\rm EW}$, the fit parameter is actually $|V_{cb}|\mathcal{F}(1)\eta_{\rm EW}$, where $\eta_{\rm EW}$ is the electroweak correction, and $\mathcal{F}(1) = h_{A_1}(1)$ is a form factor at zero-recoil. Therefore, the value of $\mathcal{F}(1)$ is required to obtain $|V_{cb}|$.
It should be noted that addition of NP effects increases the number of parameters, and experimental data alone cannot fit form factors, $V_{cb}$ and NP parameters simultaneously. 
\par 
The recent lattice QCD results from Fermilab-MILC\cite{fermilab}, JLQCD\cite{jlqcd} and HPQCD collaborations \cite{hpqcd} provided a breakthrough, by publishing the results on the form factors for this decay. Previously, only $\mathcal{F}(1)$ was known at zero recoil; but now we have results on four form factors $(f,g,\mathcal{F}_1,\mathcal{F}_2)$ in the low-recoil region. In the light of these new results, we present a simultaneous fit of Belle and lattice QCD data. Now, thanks to the lattice data, we can even fit the NP parameters, allowing us to constrain different types of NP. In addition to using the Belle binned dataset, we will also demonstrate the fit using a \cn{pseudo dataset}~\cite{emizh} – the latter will become possible once a larger amount of data becomes available at the Belle II experiment~\cite{belle2physicsbook}.  
\par 
The organisation of the article is as follows:
In Section~\ref{sec:npmodel} we introduce the effective Hamiltonian and the different NP operators relevant for {the} \BDlnu decay. 
 In Section~\ref{sec:angdist}, we write the complete angular decay distribution for the \BDlnu process in terms of the NP Wilson coefficients, hadronic and leptonic amplitudes.  
In Section~\ref{sec:lattice}, we describe the new lattice QCD data \cn{in the so-called BGL form factor parameterisation \cite{bgl}}, and show the form factor fit results to the higher recoil region using the new lattice data. 
In Section~\ref{sec:binned}, we use the Belle binned data along with the lattice data to fit form factors, $V_{cb}$ and right-handed NP Wilson coefficient, and present the fit results.
Finally, in Section~\ref{sec:unbinned}, we \cn{illustrate} the fit procedure with unbinned Belle and lattice data and perform a sensitivity study with different NP operators {and we conclude in Section~\ref{sec:conclusions}}

\section{New physics model}\label{sec:npmodel}
We consider NP particle which contributes to the semi-leptonic \BDlnu decay with $\ell=e, \mu$. 
We start with the most general effective Hamiltonian, containing all dimension six, four-fermion operators, assuming no right-handed neutrinos: 
\begin{equation}\label{eq:effective_hamiltonian}
\Heff= \frac{4G_F}{\sqrt{2}} V_{cb}\sum_{\ell=e, \mu} [ C_{ \rm  V_L}^\ell O^\ell_{ \rm V_L} + C_{\rm  V_R}^\ell O^\ell_{ \rm V_R} + {C_{\rm  S}^\ell O^\ell_{\rm  S}}+C_{\rm  P}^\ell O^\ell_{\rm  P} + C_{\rm  T}^\ell O^\ell_{\rm  T}]
\end{equation}
where ${\rm V_L, V_R, S, P, T}$ indicate the left-handed, right-handed, scalar, pseudoscalar, and tensor terms, respectively.  The corresponding operators are given as
\begin{align} 
\begin{aligned}
O^{\ell}_{\rm V_L} &= (\bar{c}_L \gamma^{\mu} b_L)(\bar{{\ell}}_L \gamma_{\mu} \nu_{{\ell}L}) \\
O^{\ell}_{\rm V_R} &= (\bar{c}_R \gamma^{\mu} b_R)(\bar{{\ell}}_L \gamma_{\mu} \nu_{{\ell}L}) \\
O^{\ell}_{\rm S} &=  (\bar{c}  b)(\bar{{\ell}}_R  \nu_{{\ell}L}) \\
O^{\ell}_{\rm P} &= (\bar{c} \gamma^5 b)(\bar{{\ell}}_R  \nu_{{\ell}L}) \\
O^{\ell}_{\rm T} &= (\bar{c}_R \sigma^{\mu\nu} b_L)(\bar{{\ell}}_R \sigma_{\mu\nu}  \nu_{{\ell}L}) 
\end{aligned}
\end{align}
where $\psi_{L/R}=\frac{1\mp \gamma_5}{2}\psi$ with $\psi=\{ c,  b,  \ell,  \nu \}$. {The contribution from the scalar operator $O^{\ell}_{\rm S}$ does not appear in the following}. 
We ignore the superscript $\ell$ in the following discussion.
In the SM, $C_{\rm V_L} = 1$, while rest of the Wilson coefficients are null.

\section{Angular distribution of the \BDlnu decay}\label{sec:angdist}
\begin{figure}[t]
\setlength{\unitlength}{1mm}
  \centering
  \begin{picture}(140,60)
    \put(0,-1){
      \includegraphics*[width=140mm]{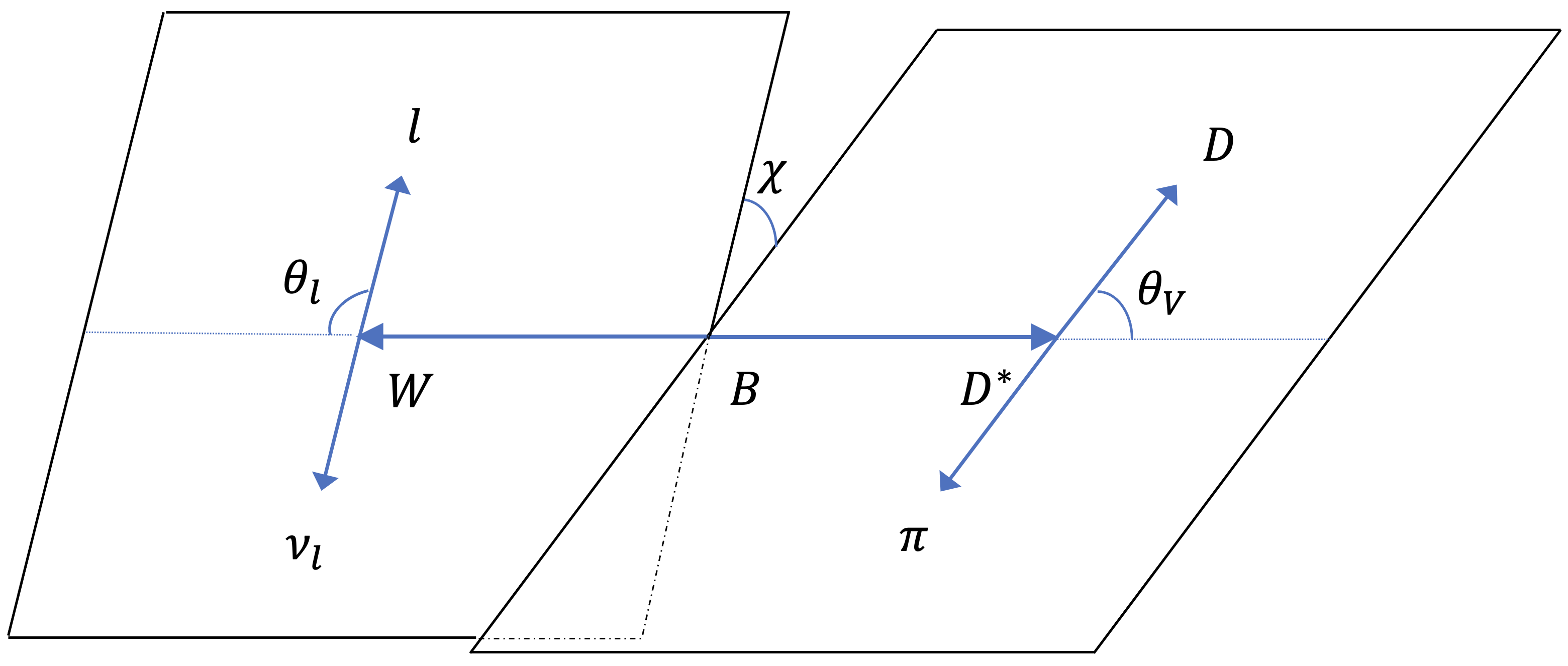}
    }
  \end{picture}
  \caption{\small Decay angles for the \BDlnu decay, where  
$\theta_V$ is the angle between the direction of $D$ meson and the direction opposite to that of $B$ meson, in the $D^*$ meson rest frame.   Similarly, 
$\theta_\ell$ is the angle between the direction of lepton and the direction opposite to that of $B$ meson, in the virtual $W$ rest frame.  $\chi$ is the angle between the two decay planes.}
\label{fig:decayangles}
\end{figure}
Using the Hamiltonian of Eq.~\eqref{eq:effective_hamiltonian}, we compute the angular distribution of the \BDlnu decay: 
\begin{align}
\begin{aligned}
\frac{\text{d} \Gamma( \bar B \to D^{*} (\to D \pi) \, \ell^- \, \bar \nu_\ell)}{\text{d} q^2 \, \text{d}\cos\theta_V \, \text{d}\cos\theta_{\ell} \, \text{d}\chi} 
= &\frac{3}{8(4\pi)^4}  \frac{(q^2 - m_\ell^2)^2 |\vec{p}_{D^*}|^2}{M_B^2 q^2} G_F^2 |V_{cb}|^2 |\eta_{\rm EW}|^2 \mathcal{B}(D^*\rightarrow D \pi)  \\
&\sum_{\lambda_{\ell}, \lambda_{\nu_{\ell}}} |\Amp_{\rm V_L}+\Amp_{\rm V_R}+\Amp_{\rm P}+\Amp_{\rm T}|^2
\end{aligned}
\end{align}
\cn{where $\theta_V,\theta_\ell$ and $\chi$ are defined in Fig.~\ref{fig:decayangles} and} $q^2$ is the invariant mass of the lepton-neutrino pair. Following the method given in~\cite{korner}, we compute the amplitudes $\Amp_{\rm V_L, V_R, P, T}$. 
In this method, we introduce the $W$ boson polarisation and separate the amplitudes into the hadronic and leptonic part as: 
\begin{align}
\begin{aligned}
\Amp_{\rm V_L} &=  \, C_{\rm V_L} \sum_{\lambda_{D^*}= \pm,0} P(\lambda_{D^*}) \sum_{\lambda= t,\pm,0} g_{\lambda\lambda} H^{\lambda_{D^*}}_{\rm V_L,\lambda} L_{\lambda}^{\rm VA} \\
\Amp_{\rm V_R} &=  \, C_{\rm V_R} \sum_{\lambda_{D^*}= \pm,0} P(\lambda_{D^*}) \sum_{\lambda= t,\pm,0} g_{\lambda\lambda} H^{\lambda_{D^*}}_{\rm V_R,\lambda} L_{\lambda}^{\rm VA} \\
\Amp_{\rm P} &=  \, C_{\rm P} \sum_{\lambda_{D^*}= \pm,0} P(\lambda_{D^*}) H^{\lambda_{D^*}}_{\rm P} L^{\rm SP} \\
\Amp_{\rm T} &=  \, C_{\rm T}
\sum_{\lambda_{D^*}= \pm,0} P(\lambda_{D^*})
\sum_{\lambda= t,\pm,0} \sum_{\lambda'= t,\pm,0}g_{\lambda\lambda} g_{\lambda'\lambda'}
H^{\lambda_{D^*}}_{\rm T,\lambda \lambda'} L^{\rm T}_{\lambda \lambda'}
\end{aligned}
\end{align}
where  $P(\lambda_{D^*}) = \langle D \pi | D^* \rangle = 2 g \epsilon_{D^*}^{*\alpha}(\lambda_{D^*}) p_{D,\alpha}$ ($g$ is the $D^*$-$D$-$\pi$ coupling),  $\epsilon_{D^*}$ is the polarisation vector of $D^*$ meson and $\lambda_{D^*} \in \{ +,-,0\}$ is the polarisation of $D^*$ meson. The polarisation vectors are given in~\cite{korner}.

The hadronic {amplitudes} are given as
\begin{align}
\begin{aligned}
 H_{\rm V_L,\lambda}^{\lambda_{D^*}}(q^2)
 &= \bar \epsilon^*_\mu(\lambda) \langle D^* (p_{D^*},\epsilon_{D^*} \left(\lambda_{D^*})\right)| \bar c \gamma^\mu(1-\gamma^5) b| \bar B (p_B)\rangle\,,\\
 H_{\rm V_R,\lambda}^{\lambda_{D^*}}(q^2)
 &= \bar \epsilon^*_\mu(\lambda) \langle D^* (p_{D^*},\epsilon_{D^*} \left(\lambda_{D^*})\right)| \bar c \gamma^\mu(1+\gamma^5) b| \bar B (p_B)\rangle\,,\\
 %
 %
 H_{\rm P}^{\lambda_{D^*}}(q^2)
 &= \langle D^* (p_{D^*},\epsilon_{D^*} \left(\lambda_{D^*})\right)| \bar c \gamma^5 b| \bar B (p_B)\rangle\,,\\
 H_{\rm T,\lambda\lambda'}^{\lambda_{D^*}}(q^2)
 &= i \bar \epsilon^*_\mu(\lambda)\bar \epsilon^*_\nu(\lambda') \langle D^* (p_{D^*},\epsilon \left(\lambda_{D^*})\right)| \bar c \sigma^{\mu\nu}(1-\gamma^5) b| \bar B (p_B)\rangle,
\end{aligned}
\end{align}
and the leptonic amplitudes are given as
\begin{align}
\begin{aligned}
L^{\rm V_{L/R}}_{\lambda} &= \bar{\epsilon}_{\mu}(\lambda) \bar{u}_\ell \gamma^{\mu} (1\mp\gamma^5) v_{\bar{\nu}_{\ell}} \\
L^{\rm P} &= \bar{u}_\ell (1-\gamma^5) v_{\bar{\nu}_{\ell}} \\
L^{\rm T}_{\lambda,\lambda'} &= -i\bar{\epsilon}_{\mu}(\lambda)\bar{\epsilon}_{\nu}(\lambda') \bar{u}_\ell \sigma^{\mu\nu} (1-\gamma^5) v_{\bar{\nu}_{\ell}},
\end{aligned}
\end{align}
\cn{where} $\bar{\epsilon}$ is the polarisation vector of the $W$ boson, and $\lambda^{(')} \in \{+,-,0,t\}$ is its polarisation. Note that $\bar{\epsilon}(t)$ is not orthogonal but proportional to $W$-momentum. \cn{Again, see ~\cite{korner} for the polarisation vectors}.

Then, using the completeness relation  of the polarisation vectors of $W$ boson, we square the  hadronic and leptonic parts of the amplitudes separately. This allows us to evaluate each part in a convenient frame: we choose the $W$ rest frame for the leptonic part and $B$ rest frame for the hadronic part. 

\cn{For} hadronic part we find 
\begin{align}\label{eq:helamp}
\begin{aligned}
H_{\rm V}^\pm(q^2)  &\equiv H_{\rm V_L,\pm}^{\pm}(q^2) = -H_{\rm V_R,\mp}^{\mp}(q^2) \\
&= (M_B + M_{D^*}) A_1(q^2) \mp \frac{\sqrt{\lambda_{D^*}(q^2)}}{M_B + M_{D^*}}V(q^2) \\
H_{\rm V}^0(q^2) &\equiv H_{\rm V_L,0}^{0}(q^2) = - H_{\rm V_R,0}^{0}(q^2) \\
&= \frac{M_B + M_{D^*}}{2 M_{D^*}\sqrt{q^2}} \left[ -(M_B^2 - M_{D^*}^2 - q^2) A_1(q^2) + \frac{\lambda_{D^*}(q^2)}{(M_B + M_{D^*})^2}A_2(q^2) \right] \\
H_{\rm P}(q^2) &\equiv H_{\rm P}^0(q^2) \\
&= -\frac{\sqrt{\lambda_{D^*}(q^2)}}{M_B + m_c} A_0(q^2) \\
H_{\rm T}^\pm(q^2) &\equiv \pm H_{\rm{T},\pm t}^{\pm}(q^2) \\
&= \frac{1}{\sqrt{q^2}} \bigg[ \sqrt{\lambda_{D^*}(q^2)} T_1(q^2) \pm (M_B^2 - M_{D^*}^2) T_2(q^2) \bigg] \\
H_{\rm T}^0(q^2) &\equiv H^0_{\rm{T},+-}(q^2) = H^0_{\rm{T},0t}(q^2) \\
&= \frac{1}{2 M_{D^*}} \bigg[ (M_B^2 + 3 M_{D^*}^2 - q^2) T_2(q^2) - \frac{\lambda_{D^*}(q^2) }{(M_B^2 - M_{D^*}^2)}T_3(q^2) \bigg] \\
\end{aligned}
\end{align}
\cn{where $V$ is the vector form factor, $A_{0,1,2}$ are the axial-vector form factors and $T_{1,2,3}$ are the tensor form factors.}
These expressions agree with the results in \cite{london} \cn{and \cite{helampref2}} while we find an overall sign difference in $H_{T,0}$ with respect to~\cite{tensorff}. 

Finally the angular distribution of  $\bar{B} \rightarrow D^{*} (\rightarrow D \pi) \ell^-\bar \nu_\ell$ is found to be
\begin{align} \label{eq:rateJ}
\begin{aligned}
\frac{\text{d} \Gamma( \bar{B} \rightarrow D^{*} (\rightarrow D \pi) \ell^-\bar \nu_\ell)}{\text{d} w \text{d}\cos\theta_V \text{d}\cos\theta_{\ell} \text{d}\chi} = & \frac{6M_BM_{D^*}^2}{8(4\pi)^4}\sqrt{w^2-1}(1-2 \, w\, r+r^2)\, G_F^2 \, \left|V_{cb}\right|^2 |\eta_{\rm EW}|^2 \, \times  \\
&\mathcal{B}(D^{*} \to D \pi) \Big\{ J_{1s} \sin^2\theta_V+J_{1c}\cos^2\theta_V \\
&+(J_{2s} \sin^2\theta_V+J_{2c}\cos^2\theta_V )\cos 2\theta_\ell   \\
& +J_3 \sin^2\theta_V\sin^2\theta_\ell\cos 2\chi  \\
&  +J_4\sin 2\theta_V\sin 2\theta_\ell \cos\chi 
+J_5 \sin 2\theta_V\sin\theta_\ell\cos\chi \\ 
& +(J_{6s} \sin^2\theta_V+J_{6c}\cos^2\theta_V)\cos\theta_\ell  \\
& +J_7 \sin 2\theta_V\sin\theta_\ell \sin\chi+J_8\sin 2\theta_V \sin 2\theta_\ell\sin\chi  \\
& +J_9 \sin^2\theta_V\sin^2\theta_\ell \sin2\chi\Big\}
\end{aligned}
\end{align}
where $|{\vec{p}_{D^*}}|=M_{D^*}\sqrt{w^2-1}$ and $r\equiv \frac{M_{D^*}}{M_B}$. The $q^2$ and $w$ are related as $\sqrt{q^2}=M_{B}\sqrt{1-2wr+r^2}$.
The angular coefficients $J_i (i=1\sim 9) $ are given in Tables~\ref{table:jfunclhrh} and~\ref{table:jfuncpt}.

\begin{table}[htp]
\begin{center}
\begin{tabular}{|c||c|c|c|c|}
\hline
J & $|C_{\rm V_L}|^2 $ & $ |C_{\rm V_R}|^2$  & $ \textrm{Re}[  C_{\rm V_L} C_{\rm V_R}^*]$&$ \textrm{Im}[  C_{\rm V_L} C_{\rm V_R}^*]$ \\
\hline \hline 
$J_{1s}$ & $ \frac{3}{2} [(H_{\rm V}^+)^2 + (H_{\rm V}^-)^2] $ &  $\frac{3}{2} [(H_{\rm V}^+)^2 + (H_{\rm V}^-)^2] $ & $ - 6 H_{\rm V}^+ H_{\rm V}^-$ & 0 \\
\hline
$J_{1c}$ & $2 (H_{\rm V}^0)^2$ & $2 (H_{\rm V}^0)^2  $ & $-4 (H_{\rm V}^0)^2$ & 0\\
\hline
$J_{2s}$ & $ \frac{1}{2} [(H_{\rm V}^+)^2 + (H_{\rm V}^-)^2] $ &  $\frac{1}{2} [(H_{\rm V}^+)^2 + (H_{\rm V}^-)^2]$ & $ -2 H_{\rm V}^+ H_{\rm V}^-$ & 0 \\
\hline
$J_{2c}$ & $-2 (H_{\rm V}^0)^2$ & $-2 (H_{\rm V}^0)^2  $ & $4 (H_{\rm V}^0)^2$ & 0 \\
\hline 
$J_{3}$ & $-2 H_{\rm V}^+H_{\rm V}^- $ & $-2 H_{\rm V}^+H_{\rm V}^-  $ & $2 [(H_{\rm V}^+)^2 + (H_{\rm V}^-)^2]$ & 0 \\
\hline
$J_{4}$ & $(H_{\rm V}^+H_{\rm V}^0 + H_{\rm V}^-H_{\rm V}^0) $ & $(H_{\rm V}^+H_{\rm V}^0 + H_{\rm V}^-H_{\rm V}^0)$ & $-2(H_{\rm V}^+H_{\rm V}^0 + H_{\rm V}^-H_{\rm V}^0)$ & 0  \\
\hline
$J_{5}$ & $-2(H_{\rm V}^+H_{\rm V}^0 - H_{\rm V}^-H_{\rm V}^0) $ & $2(H_{\rm V}^+H_{\rm V}^0 - H_{\rm V}^-H_{\rm V}^0)$ & 0& 0  \\
\hline
$J_{6s}$ & $-2 [(H_{\rm V}^+)^2 - (H_{\rm V}^-)^2] $ & $2 [(H_{\rm V}^+)^2 - (H_{\rm V}^-)^2] $ & 0& 0 \\
\hline
$J_{6c}$ & 0 & 0 & 0&0  \\ 
\hline
$J_{7}$ & 0 & 0 & 0 &0 \\
\hline 
$J_{8}$ & 0 & 0 &0 & $2(H_{\rm V}^+H_{\rm V}^0 + H_{\rm V}^-H_{\rm V}^0)$  \\
\hline
$J_{9}$ & 0 & 0 & 0 & $-2[(H_{\rm V}^+)^2 - (H_{\rm V}^-)^2] $ \\
\hline
\end{tabular}
\end{center}
\caption{The angular coefficients ($J$-functions) appearing in the differential  decay rate in Eq.~\eqref{eq:rateJ}.  $H_{\rm V}^{\pm,0}$, $H_{\rm T}^{\pm,0}$ and $H_{\rm P}$ are the hadronic amplitudes, defined in Eq.~\eqref{eq:helamp}. The columns contain the left- and right-handed contributions and their interference terms (see Table~\ref{table:jfuncpt} for the pseudocalar and tensor contributions and their interference terms). The complete $J_i$-function is sum of all terms in its row (including that of Table~\ref{table:jfuncpt}), multiplied with the corresponding Wilson coefficient term in the corresponding column (written in the topmost row). }
\label{table:jfunclhrh}
\end{table}

\begin{table}
\begin{center}
\begin{tabular}{|c||c|c|c|c|}
\hline
J &  $ |C_{\rm P}|^2$ & $|C_{\rm T}|^2$ & $ \textrm{Re}[C_{\rm P} C_{\rm T}^*] $& $\textrm{Im}[C_{\rm P} C_{\rm T}^*] $\\
\hline \hline 
$J_{1s}$ & 0 & $8[(H_{\rm T}^+)^2 + (H_{\rm T}^-)^2]  $ & 0& 0 \\
\hline
$J_{1c}$ & $2 H_{\rm P}^2$ & $32 (H_{\rm T}^0)^2$ & 0 & 0\\
\hline
$J_{2s}$ & 0 & $-8[(H_{\rm T}^+)^2 + (H_{\rm T}^-)^2] $ & 0 & 0\\
\hline
$J_{2c}$ &0 & $32 (H_{\rm T}^0)^2 $ & 0 & 0\\
\hline 
$J_{3}$ &0 & $-32 H_{\rm T}^+ H_{\rm T}^-$ & 0 & 0\\
\hline
$J_{4}$ &  0 & $16 (H_{\rm T}^- H_{\rm T}^0 - H_{\rm T}^+ H_{T}^0)$ & 0 & 0\\
\hline
$J_{5}$ & 0 & 0 & $- 8 H_{\rm P} (H_{\rm T}^+ - H_{\rm T}^-)$ & 0\\
\hline
$J_{6s}$ & 0 & 0 & 0& 0 \\
\hline
$J_{6c}$ &  0 & 0 & $- 32 H_{\rm P} H_{\rm T}^0$& 0  \\ 
\hline
$J_{7}$ & 0 & 0 &  0&$8 H_{\rm P} (H_{\rm T}^+ + H_{\rm T}^-)$  \\
\hline 
$J_{8}$ & 0 & 0 & 0 &0  \\
\hline
$J_{9}$ & 0 & 0  & 0 &0 \\
\hline
\end{tabular}
\end{center}
\caption{The pseudoscalar and tensor terms of the angular coefficients in Eq.~\eqref{eq:rateJ}, as described in caption of Table~\ref{table:jfunclhrh}.}
\label{table:jfuncpt}
\end{table}

At the limit of massless leptons, the interferences occur only between left- and right-handed currents, and between pseudoscalar and tensor contributions, which means that the SM (i.e. $ C_{\rm V_L}=1$) interferes only with the right-handed NP. Since the SM does not produce any of $J_{6c, 7, 8, 9}$, a nonzero measurement of these angular coefficients indicates immediately an appearance of  NP (null-test). In particular, the $J_{7,8,9}$ are CP violating observable, which can be induced only by complex Wilson coefficients. 

As given in Eq.~\eqref{eq:helamp}, the helicity amplitudes $(H_{\rm V}^{\pm/0}, H_{\rm P}, H_{\rm T}^{\pm/0})$ are functions of form factors. Thus, the desired NP information which is engraved in the Wilson coefficients can be obtained only by knowing them. As we discuss in the next section, the vector and the axial vector form factors have recently been computed by the lattice QCD and we will use those to investigate the NP effects.

\section{The new lattice data on the form factors}\label{sec:lattice}
As mentioned in the introduction, the Belle experiment attempted to fit the form factors and the SM parameter $V_{cb}$ simultaneously. However, it is important to mention that it has always required one lattice input, such as $\mathcal{F}(1)$.  Furthermore, an inclusion of the NP effects increase the number of parameters and the experiment cannot fit all of them, as demonstrated in our previous paper \cite{emizh}: for example, in order to fit the right-handed contribution, \cn{one must} include \cn{the} form factor $h_V(1)$  as an input. Now that the lattice QCD results are available, one can perform the NP fit in a more systematic way, by including the lattice data together with the experimental data. In the following sections, we will illustrate how to perform such fit. 

{The three lattice QCD collaborations have recently published the results on the $B\to D^*$  vector and axial-vector form factors~\cite{fermilab,jlqcd,hpqcd}}. Previously only the zero-recoil results were available~\cite{fermilabold}; however, the new papers have computed the form factors on a few points around the low recoil region. As shown in the previous section, our study requires not only these two types of form factors but also the pseudoscalar and tensor ones. These form factors could in principle, be obtained from the lattice simulation as well (see e.g.~\cite{hpqcd}). On the other hand, as introduced in~\cng{\cite{tensorff}}, they can be related to the vector and axial-vector form factors. We derive these relations in Appendix~\ref{app:ffrelations}, which agree with the results in~\cite{tensorff}. Thus, in the following, we relate the pseudoscalar and tensor form factors to the vector and axial-vector form factors, for which the lattice results are available. 

In the lattice results, the form factors are parameterised using the BGL parameterisation \cite{bgl}.  Thus, in this article, we use the same parameterisation. The form factors are related to them as follows: 
\begin{align}\label{eq:bgltostdff}
\begin{aligned}
g&=\frac{2}{M_B+M_{D^*}}V \\
f&=(M_B+M_{D^*})A_1 \\
\mathcal{F}_1 &= \frac{1}{2M_{D^*}}\left[ (M_B^2-M_{D^*}^2-q^2)(M_B+M_{D^*}) A_1  -  
\frac{4M_B^2|{ \vec{p}}_{D^*}|^2}{ M_B+M_{D^*}} A_2 \right]  \\
\mathcal{F}_2 &= 2A_0 
\end{aligned}
\end{align}
where the $q^2$ (or $w$) dependence of the form factors is implicit. 
By using these definitions, the helicity amplitudes can be now written in a very simple manner: 
\begin{align} \label{helampbgl}
\begin{aligned}
H_{V}^\pm(w) &= f\mp g M_B |{\vec{p}_{D^*}}|  \\
H_{V}^0(w) &=\frac{\mathcal{F}_1 }{M_{B}\sqrt{1-2wr+r^2}} \\
H_{P}(w)&=  -\frac{\mathcal{F}_2 M_B |{ \vec{p}_{D^*}}|}{m_b + m_c} \\
H_{T}^\pm(w)&=\frac{\pm f (m_b - m_c) + g M_B |{ \vec{p}_{D^*}}|(m_b + m_c)}{ M_{B}\sqrt{1-2wr+r^2} } \\
H_{T}^0(w) &= \frac{-(m_b-m_c)(-\mathcal{F}_1 (M_B^2 - M_{D^*}^2) + 2 \mathcal{F}_2 M_B^2 |{ \vec{p}_{D^*}}|^2 ) }{(M_B^2 - M_{D^*}^2) {(M_{B}^2 + M_{D^*}^2 - 2M_{B}M_{D^*} w  )}}
\end{aligned}
\end{align}
where $m_{b(c)}$ is the $b(c)$-quark mass. Note that the pseudoscalar and tensor form factors are reduced to the above four form factors thanks to the relations given in Appendix~\ref{app:ffrelations}.
\par
The momentum dependence of these form factors are given \cn{in $z-$ expansion, where}
\begin{equation}
z \equiv \frac{\sqrt{w+1}-\sqrt{2}}{\sqrt{w+1}+\sqrt{2}},
\end{equation}
\cn{and the form factors are}
\begin{align} \label{eq:bglff}
\begin{aligned}
g(z) &= \frac{1}{P_{1^-}(z)\phi_g(z)} \sum_{n=0}^{\infty} a^g_n z_n,  \quad \quad \;
f(z) = \frac{1}{P_{1^+}(z)\phi_f(z)} \sum_{n=0}^{\infty} a^f_n z^n, \\
\mathcal{F}_1(z) &= \frac{1}{P_{1^+}(z)\phi_{\mathcal{F}_1}(z)} \sum_{n=0}^{\infty} a^{\mathcal{F}_1}_n z_n,  \quad
\mathcal{F}_2(z) = \frac{1}{P_{0^-}(z)\phi_{\mathcal{F}_2}(z)} \sum_{n=0}^{\infty} a^{\mathcal{F}_2}_n z_n .
\end{aligned}
\end{align}
Note that the expansion coefficients must satisfy the unitarity conditions: 
\begin{equation}
\sum_{n=0}^\infty  (a^g_n)^2<1, \quad 
\sum_{n=0}^\infty  (a^f_n)^2+(a^{\mathcal{F}_1}_n)^2<1, \quad 
\sum_{n=0}^\infty  (a^{\mathcal{F}_2}_n)^2<1 \label{unitarity}
\end{equation}
From Eq.~\eqref{eq:bgltostdff}, we can see that the form factors are not completely independent.

\begin{table}[t]
\begin{center}
\begin{tabular}{|c||c|c|}
\hline 
Type & $B_c^{(*)}$ mass (GeV) & $\chi^{T, L}_{1^{\pm}}$ \\ \hline\hline
$f$, $\mathcal{F}_1$ & 6.739, 6.750, 7.145, 7.150  & $3.894\times 10^{-4}\  {\rm GeV}^{-2}$\\ \hline
$g$ & 6.329, 6.920, 7.020, 7.280 &$5.131\times 10^{-4}\  {\rm GeV}^{-2} $ \\ \hline
$\mathcal{F}_2$ & 6.275, 6.842, 7.250 &$1.9421\times 10^{-2}$  \\\hline
\end{tabular}
\end{center}
\caption{The input parameters for \cn{$P_{1^-, 1^+, 0^-}$ and $\phi_{g, f, \mathcal{F}_1, \mathcal{F}_2}$ functions} used in the BGL form factors given in Eq.~\eqref{eq:bglff}. We \cn{take} the values \cn{used}  in~\cite{fermilab}}
\label{table:bglinput}
\end{table}

At the zero-recoil limit, i.e. $|{ \vec{p}_{D^*}}|=0$ $(w_{\rm min}=1)$, we find that both $f$ and $\mathcal{F}_1$ are written only by the $A_1$ form factor and are related as  
\begin{equation}
\mathcal{F}_1(0)=(M_B-M_{D^*})f(0) \label{eq:relF1}
\end{equation}
On the other hand, at the maximum recoil, $q^2=0$ $(w_{\rm max}=\frac{M_B^2+M_{D^*}^2}{2M_BM_{D^*}})$, we have a condition, $A_0=\frac{M_B+M_{D^*}}{2M_{D^*}}A_1-\frac{M_B-M_{D^*}}{2M_{D^*}}A_2$, to avoid \cn{the $q^2$ pole in the definition of axial-vector operator matrix element \cite{tensorff}}. This leads to 
\cn{
\begin{align}
\mathcal{F}_1(z(w_{\rm max}))=\frac{(M_B^2-M_{D^*}^2)}{2}\mathcal{F}_2(z(w_{\rm max})) \label{eq:relF2}
\end{align}
}
\cn{Taking into account these relations, we can eliminate two expansion parameters. We choose the lowest order $\mathcal{F}_1$ constant, i.e $a^{\mathcal{F}_1}_0$ and the highest order $\mathcal{F}_2$ constant $a^{\mathcal{F}_2}_{j_{\rm max}}$ to eliminate.}

The functions $P_{1^-, 1^+, 0^-}$ and $\phi_{g, f, \mathcal{F}_1, \mathcal{F}_2}$, whose expressions can be found in the original article~\cite{bgl}, contain several input parameters. We follow the reference~\cite{fermilab} (originally obtained in~\cite{gambino}) and use the values given in Table~\ref{table:bglinput} as input parameters\footnote{These values are very different from the ones used in \cng{ \cite{belle19} } which actually picks the values from \cite{bgl}}, with $n_I=2.6$, $M_B = 5.280$GeV and $M_{D^*} = 2.010$GeV.

\subsection{Lattice QCD data and their extrapolations to higher $w$ regions}
\begin{table}[t]
\begin{center}
\begin{tabular}{|c || d{9} | d{9} | d{9} | d{9} |}
\hline
\multicolumn{1}{|c||}{Form factor} & \multicolumn{1}{c|}{JLQCD} & \multicolumn{1}{c|}{FM } & \multicolumn{1}{c|}{JLQCD (prior)} & \multicolumn{1}{c|}{FM (prior)} \\
\hline\hline
$a^f_0$ &  0.01197(19) & 0.01209(19)  & 0.01197(19)  & 0.01208(19)  \\
$a^f_1$ &  0.020(11) & -0.012(20)  & 0.021(10)  &  0.001(15) \\
$a^f_2$ &  0.00(49) & 0.8(14)  & -0.08(43)  & 0.04(79)  \\
\hline
$a^g_0$ &  0.0294(18) & 0.0329(12)  & 0.0293(18) &  0.0328(12) \\
$a^g_1$ & -0.057(51)  & -0.15(10) & -0.057(37)  & -0.136(55) \\
$a^g_2$ &  1.0(31) &  0.9(55) &  0.72(95) &  0.27(98) \\
\hline
$a^{\mathcal{F}_1}_1$ & 0.0010(42)  & -0.0080(38) &  0.0012(41) & -0.006(3) \\
$a^{\mathcal{F}_1}_2$ & 0.04(21)  & 0.14(23)  &  0.03(20) & 0.01 (15) \\
\hline
$a^{\mathcal{F}_2}_0$ & 0.0484(17)  & 0.0500(15)  & 0.0485(17)  & 0.0499 (15) \\
$a^{\mathcal{F}_2}_1$ &  -0.055(89) & -0.377(89)  & -0.054(88) & -0.324 (77) \\
\hline
\end{tabular}
\end{center}
\caption{The BGL form factor fitted values with the lattice data from JLQCD and Fermilab-MILC (FM) collaborations. {The 2nd-3rd and 4th-5th columns contain results of fit respectively without and with  the unitarity condition, using the prior given in Eq.~\eqref{eq:prior}.} }
\label{table:latticeff}
\end{table}

Lattice QCD can compute the values of the form factors in the low $w$ region. 
Fermilab-MILC (FM) and JLQCD collaborations give the values of the four form factors, $F\equiv(f, g, \mathcal{F}_1, \mathcal{F}_2)$, at these three points:  $w_{\rm FM}=(1.03. 1.10, 1.17)$ for Fermilab-MILC collaboration and $w_{\rm JLQCD}=(1.025. 1.060, 1.100)$ for JLQCD~\footnote{{Hereafter, we focus only on the results provided by these two collaboration as the  result from~\cite{hpqcd} is given in terms of the HQET form factors and requires an extra transformation.}}. They also provides  the $12\times 12$ covariance matrix for them.  Using these lattice data, we fit the BGL parameters, $\vec{a}_{\rm BGL}=(a^f_n, a^g_n, a^{\mathcal{F}_1}_n, a^{\mathcal{F}_2}_n)$, given in Eq.~\eqref{eq:bglff}. The $\chi^2$ can be symbolically written as 
\begin{equation}\label{eq:chi2lattice}
\chi^2_{\rm latt}(\vec{a}_{\rm BGL})=(\vec{F}_{\rm BGL}(w_{\rm latt})-\vec{F}_{\rm latt}(w_{\rm latt}))V_{\rm latt}^{-1}(\vec{F}_{\rm BGL}(w_{\rm latt})-\vec{F}_{\rm latt}(w_{\rm latt}))^T,
\end{equation}
where ''latt'' is either FM or JLQCD.  The $\vec{F}_{\rm latt}(w_{\rm latt})$ are the form factors evaluated by the lattice and $V_{\rm latt}$ is the covariance matrix mentioned above. The $\vec{F}_{\rm BGL}(w_{\rm latt})$ are the form factor expanded in terms of the BGL parameters $\vec{a}_{\rm BGL}$ and evaluated, using  Eq.~\eqref{eq:bglff}, at $w_{\rm latt}$. We keep terms up to $n=2$ in the following. Using the two relations in Eqs.~\eqref{eq:relF1} and \eqref{eq:relF2}, we eliminate $a^{\mathcal{F}_1}_0$ and $a^{\mathcal{F}_2}_2$ and we fit 10 parameters.  The fit results for JLQCD and Fermilab-MILC are given in Table~\ref{table:latticeff}, in the second and third columns, respectively. 
\begin{figure}[t!]
\begin{center}
\begin{tabular}{cc}
\includegraphics[scale=0.42]{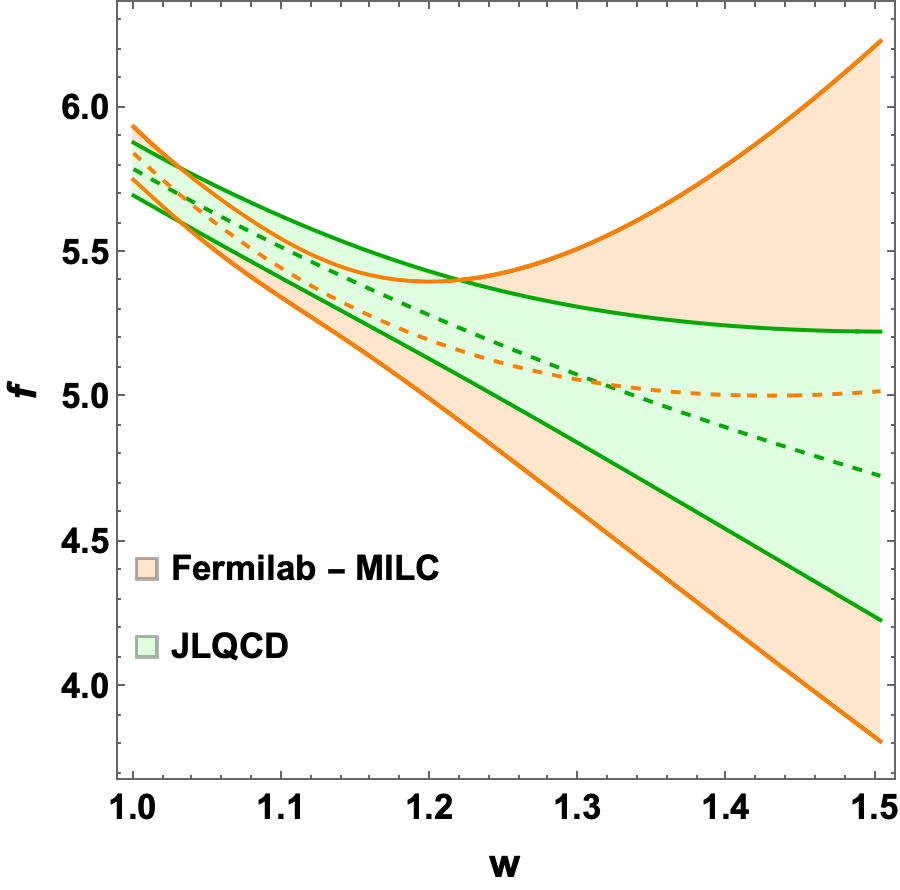} & 
\includegraphics[scale=0.43]{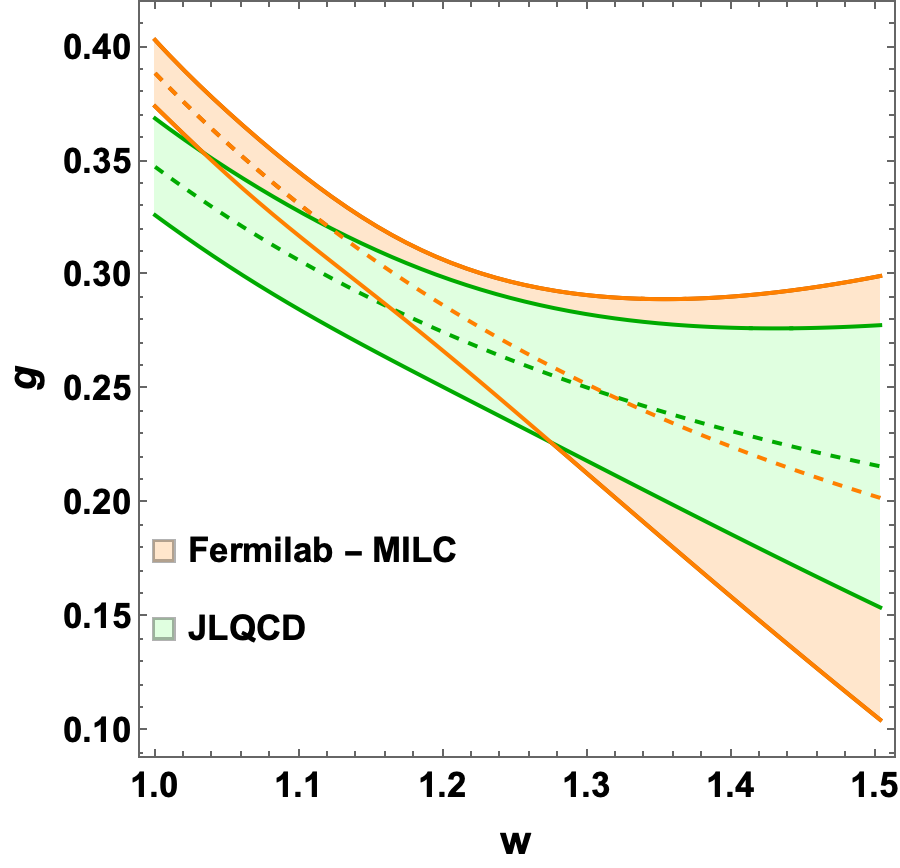} \\
\includegraphics[scale=0.42]{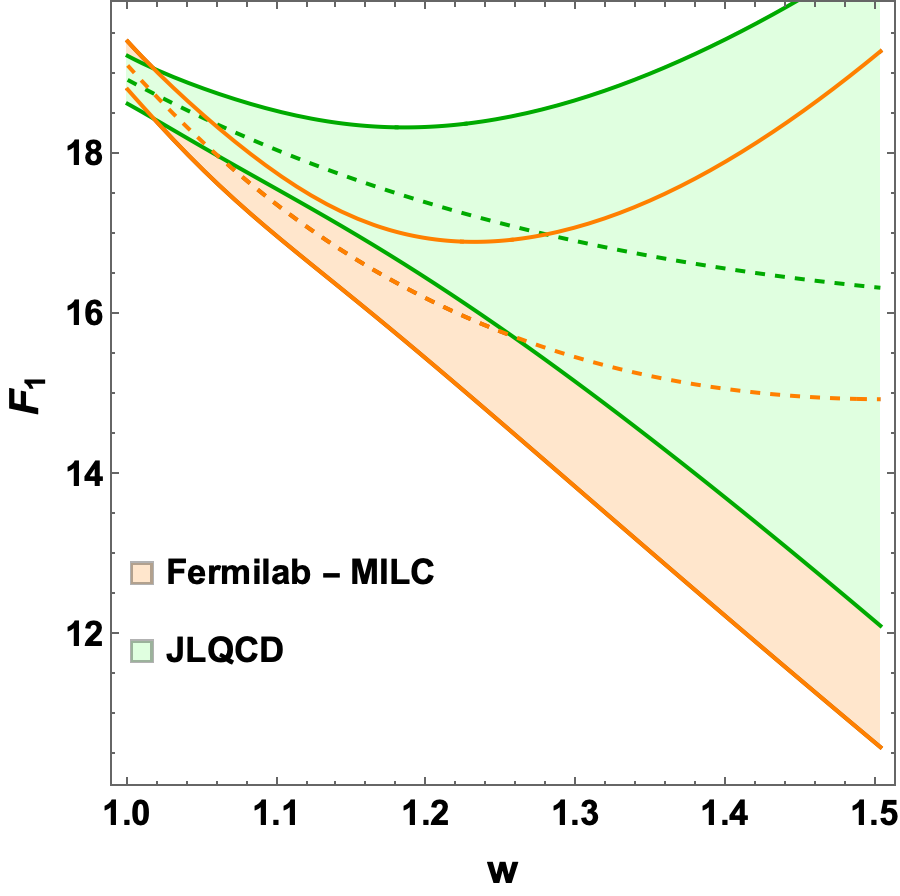} & 
\includegraphics[scale=0.43]{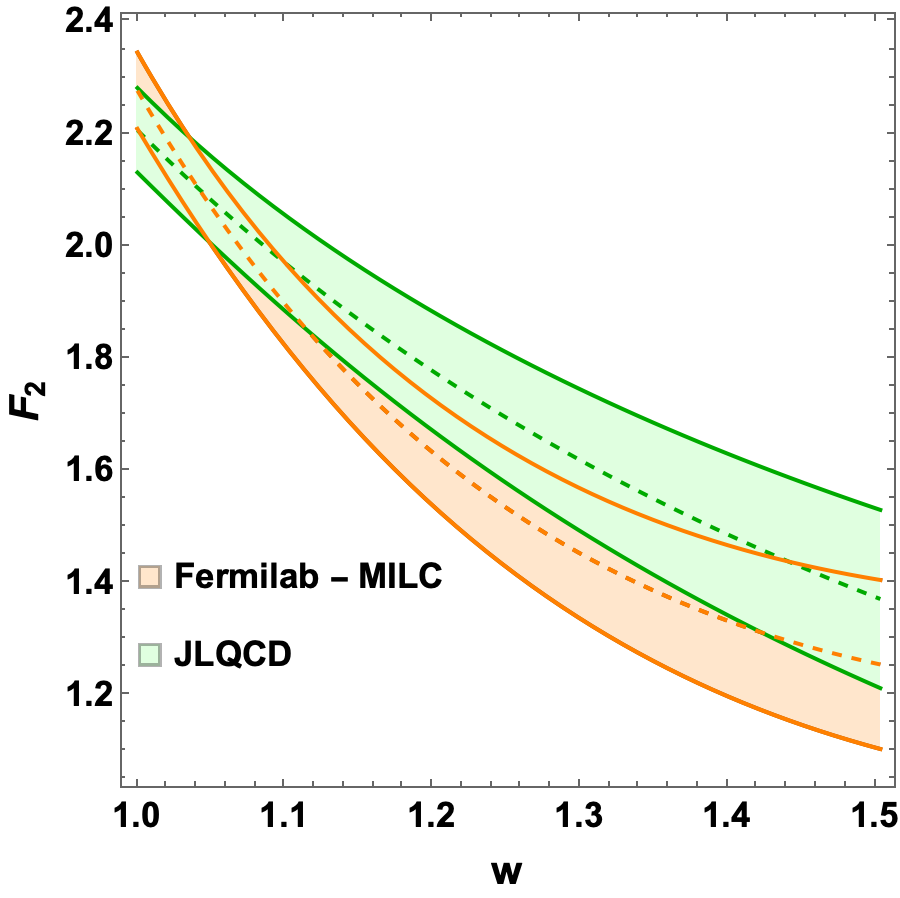}
\end{tabular}
\end{center}
\caption{{The extrapolated $w$-dependence of the form factors using the BGL parameterisataion: the green and orange bounds are obtained by using the lattice QCD inputs at the low $w$ region by the JLQCD and the Fermilab-MILC collaborations,  respectively. No unitarity condition is imposed in these plots.} }
\label{fig:latticeff}
\end{figure}

Previously, the only lattice QCD result available was for $\mathcal{F}(w)$ at zero-recoil: 
\begin{equation}\label{eq:latticezerorecoil}
\mathcal{F}(1) \equiv \frac{1}{2\sqrt{M_B M_{D^*} }} \frac{a^f_0}{P_{1^+}(0)\phi_f(0)} 
\end{equation}
with the value being {$\mathcal{F}(1) = 0.906 \pm 0.004 \pm 0.012$~\cite{fermilabold}}.  With the new lattice results, the fitted value of  $a^f_0$ above gives: 
\[\mathcal{F}^{\rm JLQCD}(1)=0.887\pm 0.014, \quad \mathcal{F}^{\rm FM}(1)=0.896\pm 0.014\]

Using these values, we can extrapolate the form factors to higher value of $w$. The extrapolated plots are shown in Fig.~\ref{fig:latticeff}. As mentioned earlier,  the expansion parameters have constraints coming from the unitarity, i.e. Eq.~(\ref{unitarity}). \cn{Our understanding is that an additional constraint is sometimes added, which is given as follows:}
\begin{equation}\label{eq:prior}
\chi^2_{\rm unitarity-bound}=|\vec{a}_{\rm BGL}|^2.
\end{equation}

\begin{figure}[t!]
\begin{center}
\begin{tabular}{cc}
\includegraphics[scale=0.42]{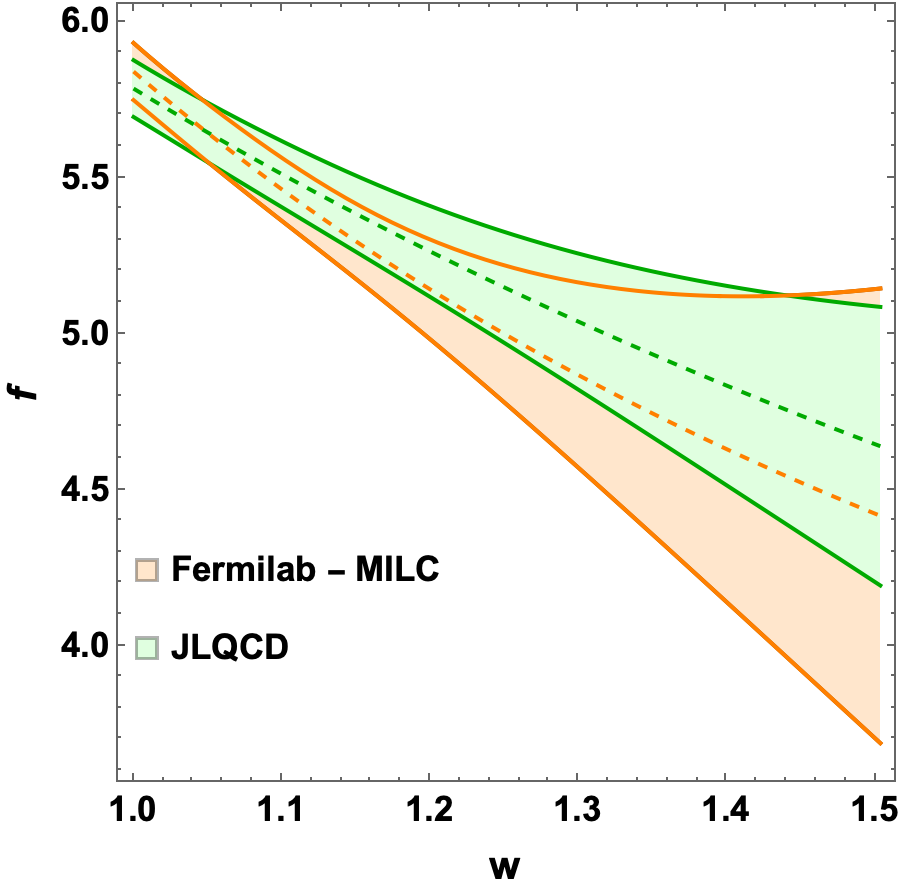} & \includegraphics[scale=0.43]{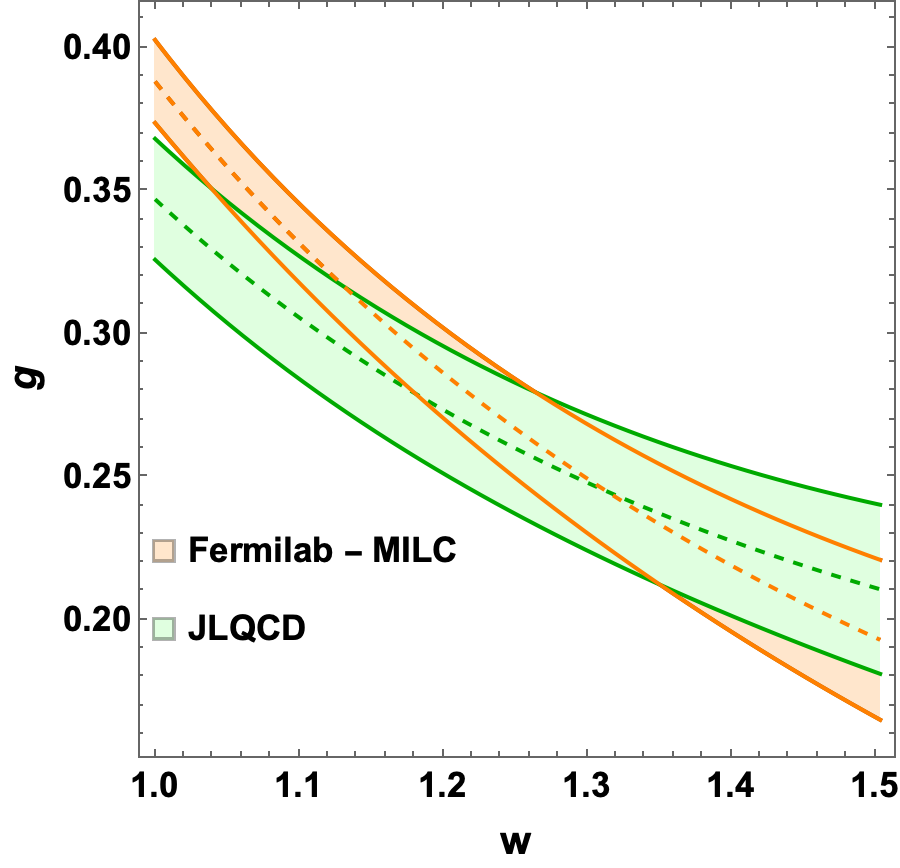} \\
\includegraphics[scale=0.42]{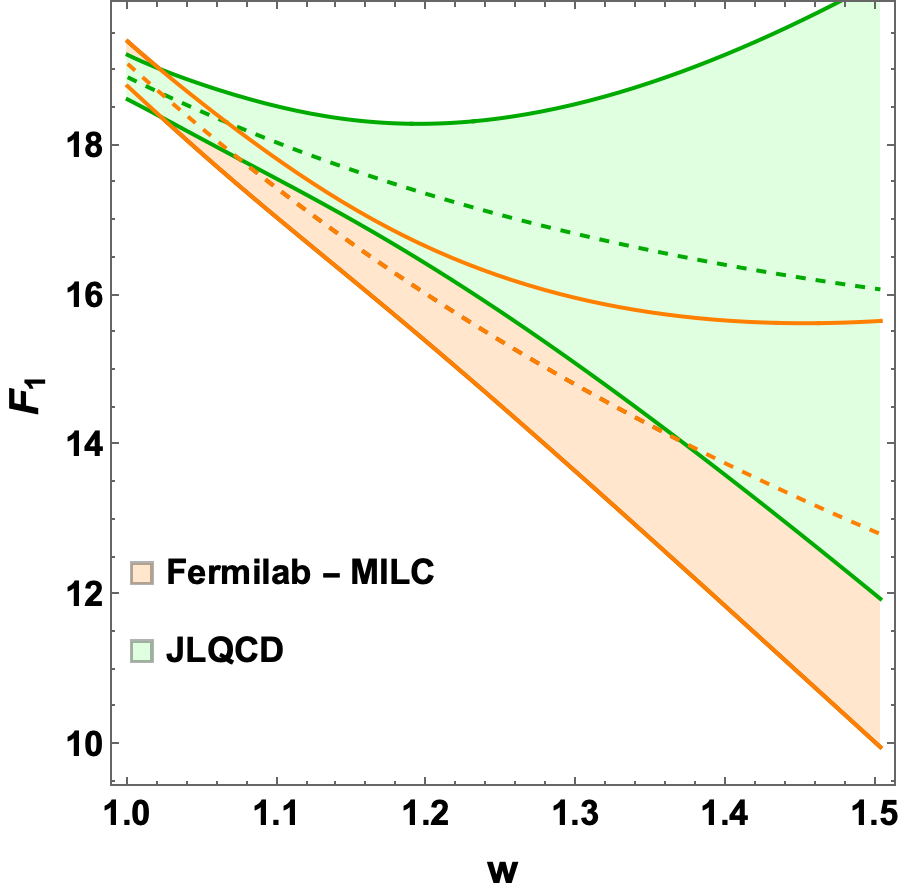} & \includegraphics[scale=0.43]{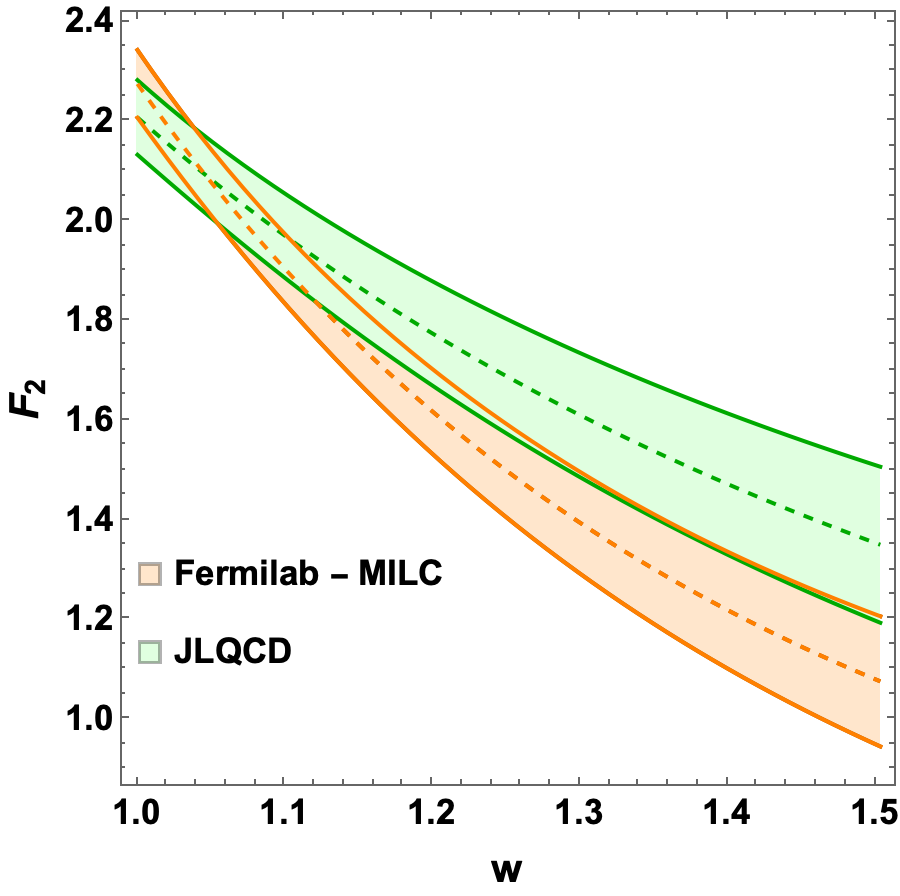} 
\end{tabular}
\end{center}
\caption{{The extrapolated $w$-dependence of the form factors using the BGL parameterisation: the green and orange bounds are obtained by using the lattice QCD inputs at the low $w$ region by the JLQCD and the Fermilab-MILC collaborations,  respectively.  The unitarity condition is imposed by using a prior given in Eq.~\eqref{eq:prior}. }}
\label{fig:latticeffprior}
\end{figure}

The results obtained with this prior are given in Table~\ref{table:latticeff} in the fourth and fifth column for JLQCD and Fermilab-MILC,  respectively. The extrapolated plots are shown in Fig.~\ref{fig:latticeffprior}. 
Even after taking into account the unitarity bound, the lattice QCD data itself does not constrain well the higher $w$ region. As we will see in the next section, the experimental data \cn{provide} a much better handle on this.

\section{The Belle binned results in the light of lattice data}\label{sec:binned}


The BaBar and Belle collaborations used their data and \cn{determined} form factors along with the CKM matrix element $V_{cb}$ in~\cite{belle17,belle19,belle23,babar04,babar07,babar19}. They use the binned distribution of one of the four kinematical variables, $w, \cos\theta_l, \cos\theta_V$ or $\chi$. \cn{Each one} is a one dimensional distribution where the other 3 variables are integrated \cn{out}: 
\begin{align}
\begin{aligned}
d\Gamma(w)&= \frac{8K}{9}  \pi \Big\{ 6J_{1s} + 3J_{1c} - 2J_{2s} - J_{2c} \Big\}  \\
d\Gamma(\cos\theta_V)&= -\frac{4K}{3}  \pi \Big\{
{-\frac{8}{3}J_{1s}}+({\frac{8}{3}J_{1s}-4J_{1c}})\cos^2\theta_V \Big\}  \\
d\Gamma(\cos\theta_\ell )&= \frac{4K}{3} \pi \Big\{ ({\frac{4}{3}J_{1s}+2J_{1c}}) +{2J_{6s}}\cos\theta_\ell +({\frac{4}{3}J_{1s}-2J_{1c}})\cos^2\theta_\ell\Big\} \\
d\Gamma(\chi)&= \frac{4K}{9}\Big\{ ({\frac{16}{3}J_{1s}+4J_{1c}})+{4 J_{3}} \cos (2 {\chi})+{4  J_{9}} \sin (2 {\chi}) \Big\} 
\end{aligned}
\end{align}
where $K = \frac{3M_BM_{D^*}^2}{4(4\pi)^4}\sqrt{w^2-1}(1-2 \, w\, r+r^2)\, G_F^2 \, {|\eta_{\rm EW}|^2}\left|V_{cb}\right|^2  \mathcal{B}(D^{*} \to D \pi)$.
They divide each distribution into 10 bins and the number of events for the total of 40 bins are published together with the detector response matrix and the selection efficiency, the systematic uncertainties and the $40\times 40$ statistical uncertainty correlation matrix. {The correlation among the bins are obtained using the SM theoretical formula. }

{In this section, we first reproduce the result of Belle~\cite{belle19}, where the form factors as well as $|V_{cb}|$ are extracted from the data within the SM framework. As mentioned earlier, $|V_{cb}|$ is an overall factor that cannot be extracted without an input of at least one form factor. In Section~\ref{sec:binned_ff}, we first obtain the {\it tilde} parameters,  $\tilde{a}_{\rm BGL}\equiv a_{\rm BGL}|V_{cb}|\eta_{EW}$, which is independent of the form factor. Then, in Section~\ref{sec:binned_smfit}, we use the lattice results introduced in the Section~\ref{sec:lattice}, and fit the form factors and $|V_{cb}|$ simultaneously. In Section~\ref{sec:binned_npfit}, we extend our model from SM to the one including right-handed current, parameterised by $C_{\rm V_R}$ and perform the fit of form factor, $|V_{cb}|$ and $C_{\rm V_R}$. There is a caveat on the last analysis: the correlation of the Belle result is obtained based on the SM theory formula. Thus, the results of Section~\ref{sec:binned_npfit} \cn{are just meant to provide a proof of principle} and we suggest the Belle collaboration to \cn{verify whether or not} the correlations including right-handed model \cn{depart significantly from the one obtained with the SM}. }

\subsection{Form factor determination with Belle data}\label{sec:binned_ff}

{In this section we  obtain the {\it tilde} parameters from the Belle data published in~\cite{belle19}. The $\chi^2$ to minimise can be written as
\begin{align}\label{eq:chi2belle}
\chi^2_{\rm Belle}(\vec{\tilde{a}}_{\rm BGL})= \sum_{i,j}(N_i^{\rm obs}-N_i^{\rm exp})C^{-1}_{ij}(N_j^{\rm obs}-N_j^{\rm exp}) 
\end{align}
where $N_i^{\rm obs}$ is the observed number of events, $N^{\rm exp}_i$ is the expected number of events and $C_{ij}$ is the covariance matrix. To avoid the problem with the highly correlated experimental uncertainties, we follow the procedure given in~\cite{bellefittechnique} and perform a Toy Monte Carlo (MC): the MC samples are generated by using the Cholesky decomposition method for both statistical and systematic covariance matrices $C_{\rm stat, syst}$
\begin{equation} 
N^{{\rm obs}\prime}=N^{\rm obs} +Lu, \quad C=LL^T
\end{equation}
where $u$ is the random value taken from a Gaussian distribution of mean 0 and variance 1. We iterate the fit a few thousands times with this MC sample and average the result to obtain the best fit values for the parameters as well as the error matrix. }

\begin{table}
\begin{center}
\begin{tabular}{|c || d{13} |}
\hline
\multicolumn{1}{|c||}{Form factor} & \multicolumn{1}{c|}{Value}  \\
\hline\hline
$ \tilde{a}^f_0 $ & 0.000459(9)(20) \\ 
$ \tilde{a}^f_1 $ & 0.0017(13)(13)\\
$\tilde{a}^f_2 $ &-0.013(43)(32) \\
\hline 
$\tilde{a}^g_0$ & 0.0016(6)(5)\\
$\tilde{a}^g_1$ & -0.019(24)(21)\\
$\tilde{a}^g_2 $ & -0.02(17)(17)\\
\hline 
$\tilde{a}^{\mathcal{F}_1}_1$ & -0.0003(1)(1)\\
$ \tilde{a}^{\mathcal{F}_1}_2$ & -0.0028(19)(16) \\
\hline
\end{tabular}
\end{center}
\caption{The BGL parameters fitted by using the Belle data published in \cite{belle19}. The tilde notation is defined as $\tilde{a}_{\rm BGL}\equiv a_{\rm BGL}|V_{cb}|\eta_{\rm EW}$. The first error is statistical and the second is systematic.}
\label{table: belleff}
\end{table}
\begin{figure}
\begin{center}
\begin{tabular}{cc }
\includegraphics[scale=0.5]{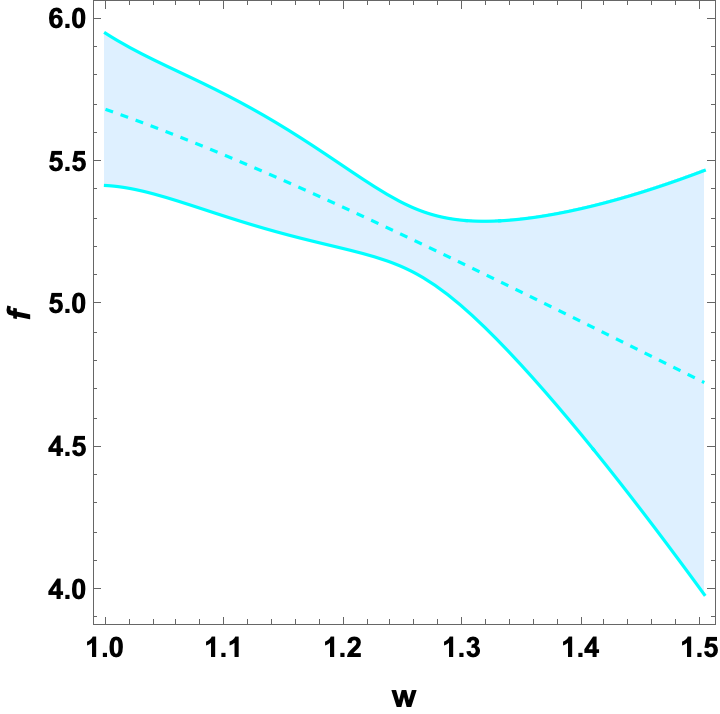} & \includegraphics[scale=0.5]{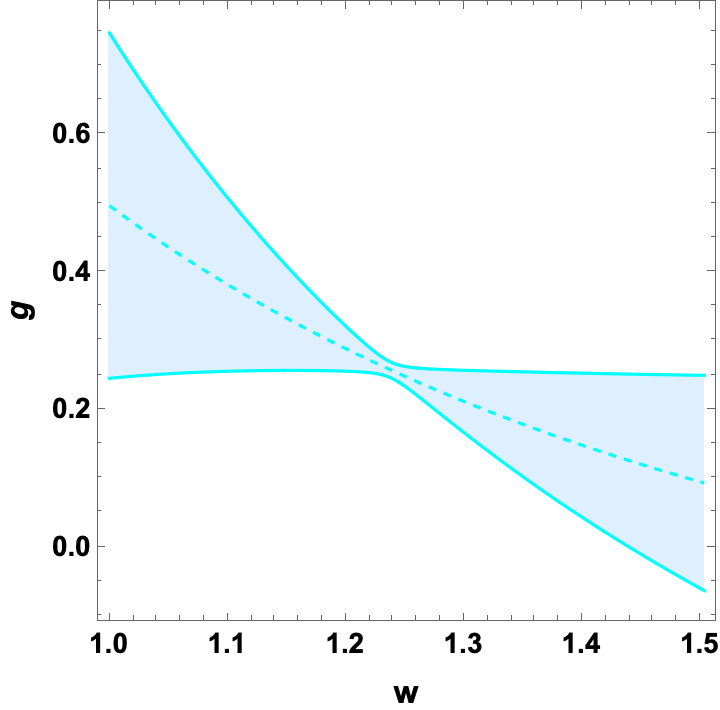} \\
\multicolumn{2}{c}{\includegraphics[scale=0.5]{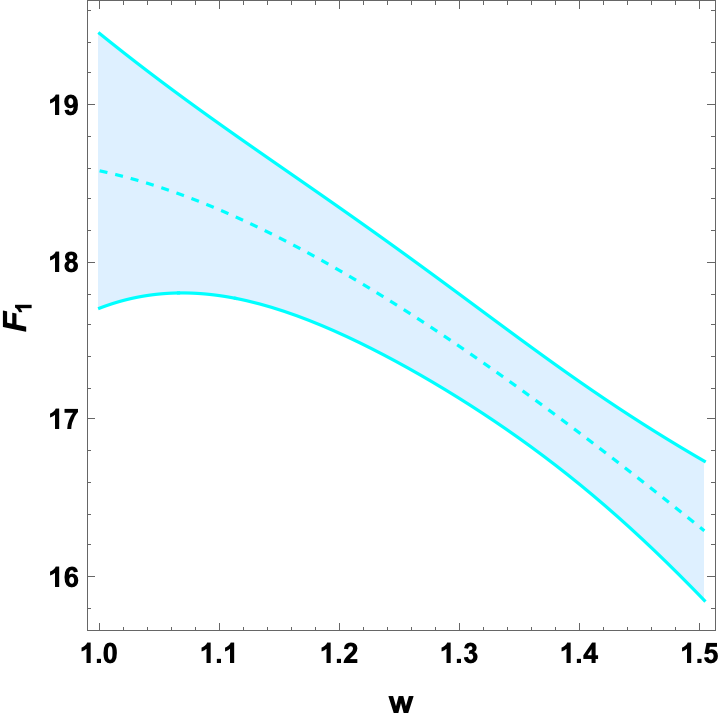}} 
\end{tabular}
\end{center}
\caption{{The $w$-dependence of the form factors using the fitted BGL parameters with the Belle data ~\cite{belle19}. Note that the originally obtained BGL parameters are in the {\it tilde} form, i.e. $\tilde{a}_{\rm BGL}\equiv a_{\rm BGL}|V_{cb}|\eta_{\rm EW}$. In this figure, we use a factor $|V_{cb}|\eta_{\rm EW}=0.039$ to relate it to the BGL parameter, $a_{\rm BGL}$ so as to  be able to compare to the $w$-dependence of the form factors in the other figures.}}
\label{fig:belleff}
\end{figure}

{The obtained results are given in Table~\ref{fig:belleff}, where the first error is statistical and the second one is systematic. Since the distribution doesn't depend on $\mathcal{F}_2$, we fit only eight parameters here. The result agrees qualitatively well with the results given in~\cite{belle19}. Using this results, we may draw the $f, g, F_{1}$ form factors in terms of $w$. However, as the obtained result from the fit is $\tilde{a}_{\rm BGL}$, we have to scale it by a factor $|V_{cb}|\eta_{\rm EW}$ to obtain ${a}_{\rm BGL}$. Just for the sake of comparing the $w$ dependence with Fig.~\ref{fig:latticeff}, we chose here \cn{$|V_{cb}|\eta_{\rm EW}=0.039$} and the result is given in Fig.~\ref{fig:belleff}.}

\subsection{SM fit of the Belle data with the new lattice data }\label{sec:binned_smfit}
\begin{figure}
\begin{center}
\begin{tabular}{cc}
 \includegraphics[scale=0.42]{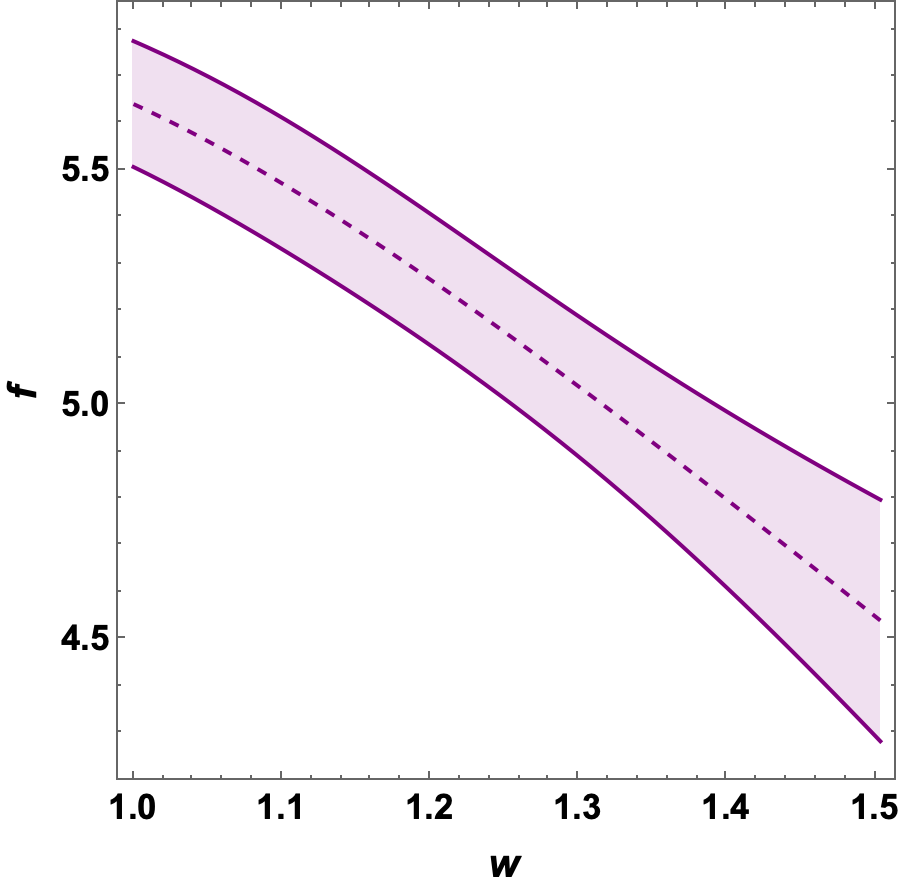} & 
 \includegraphics[scale=0.43]{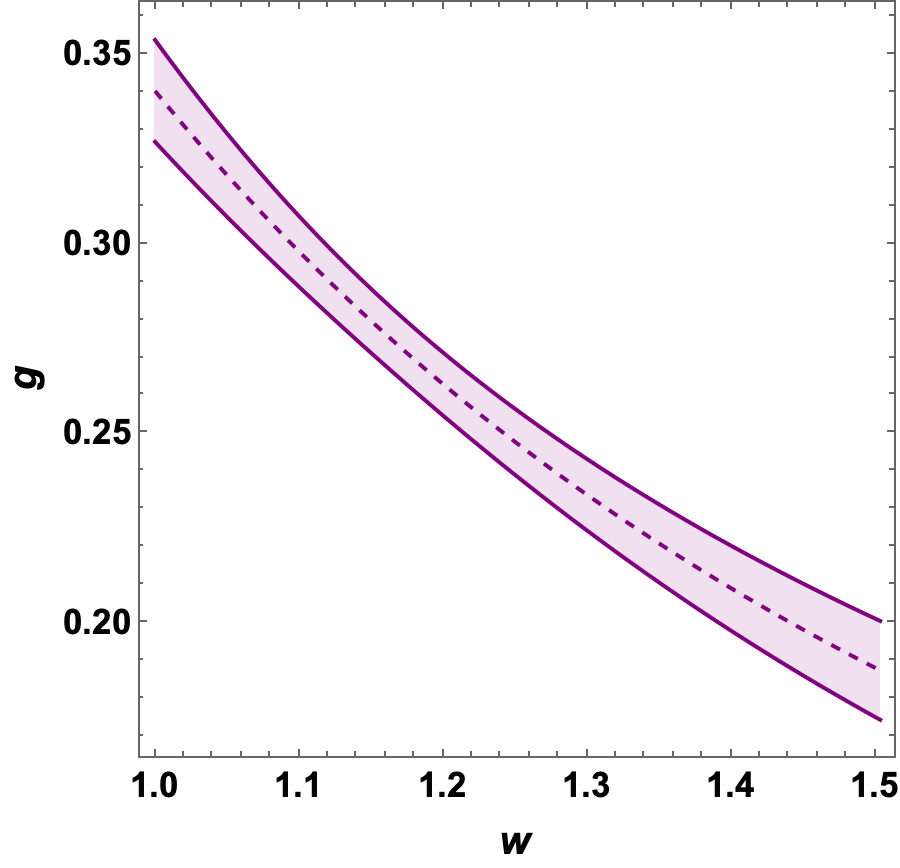} \\
 \includegraphics[scale=0.42]{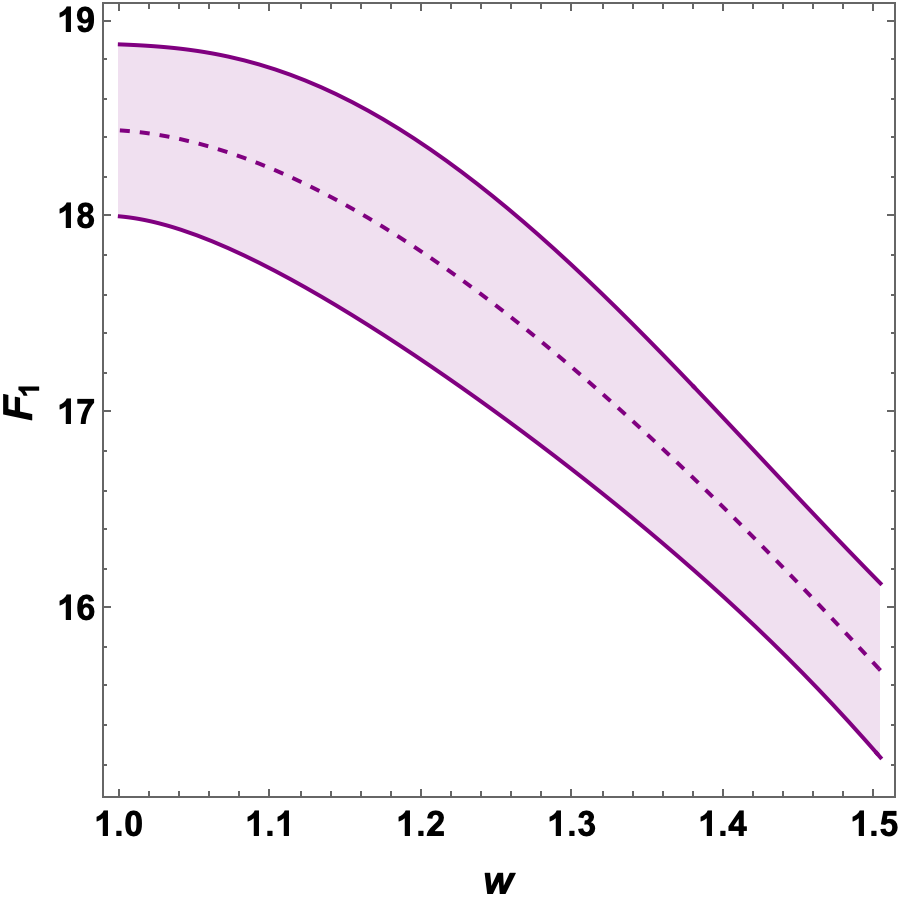} & \includegraphics[scale=0.43]{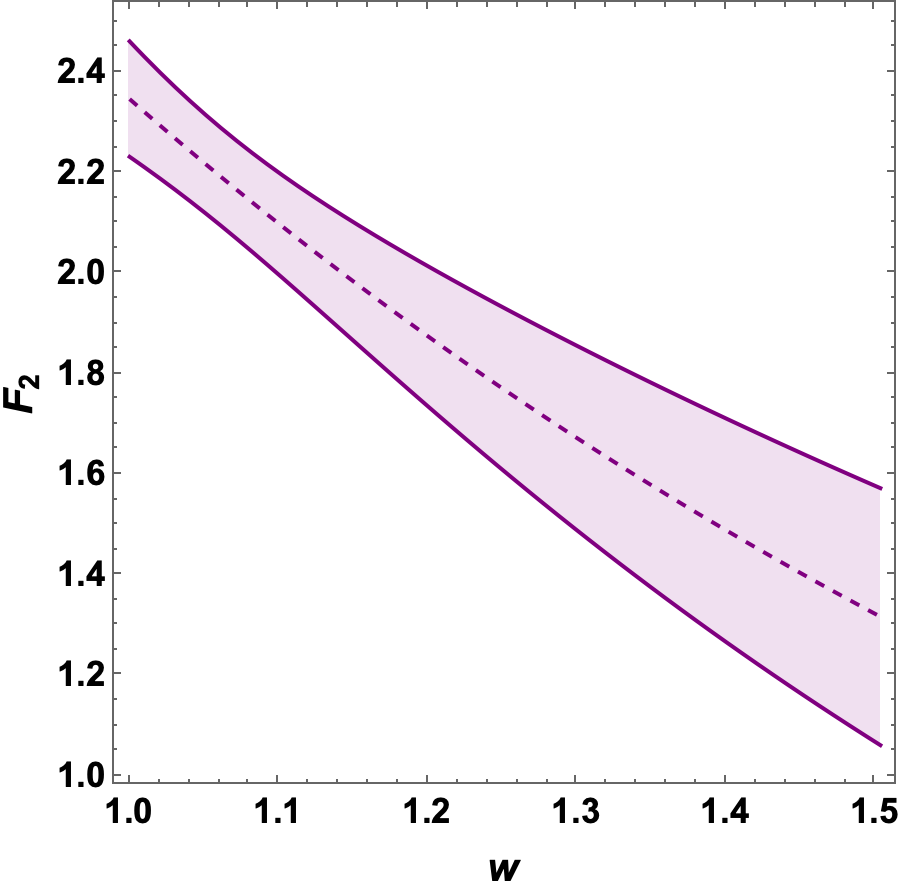} 
\end{tabular}
\end{center}
\caption{{The $w$-dependence of the form factors using the fitted BGL parameters with the Belle data ~\cite{belle19} and the JLQCD lattice data~\cite{jlqcd}.}}
\label{fig:bellelatticeff_jlqcd}
\end{figure}

\begin{figure}
\begin{center}
\begin{tabular}{cc}
\includegraphics[scale=0.42]{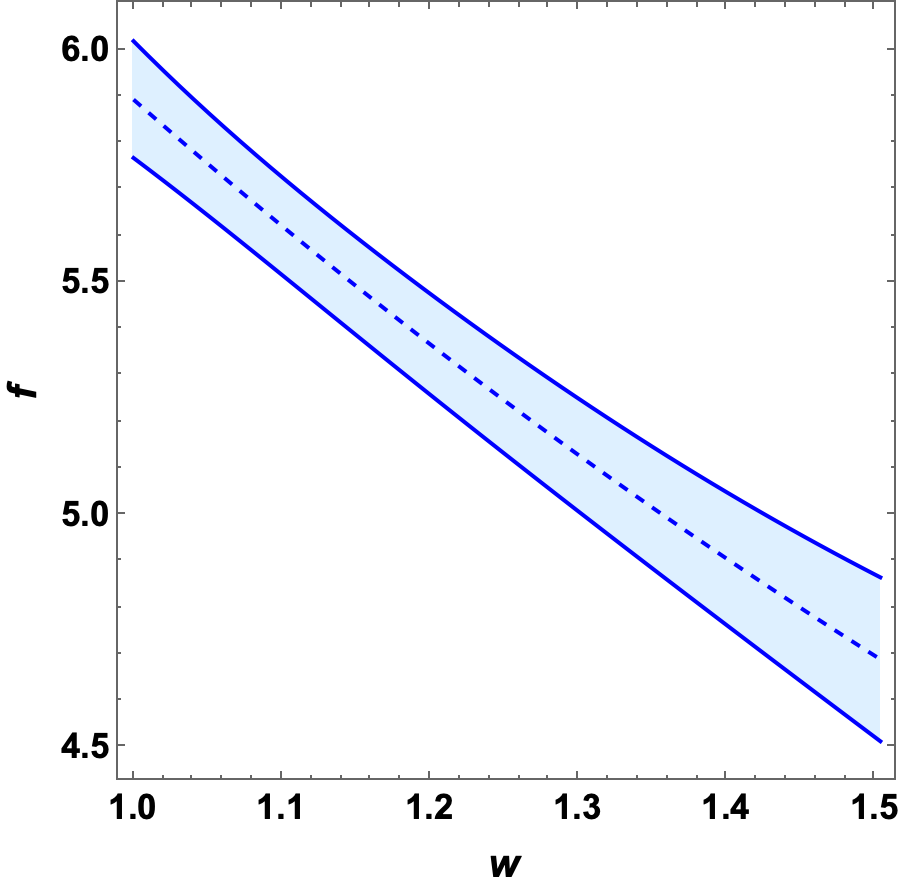} & \includegraphics[scale=0.43]{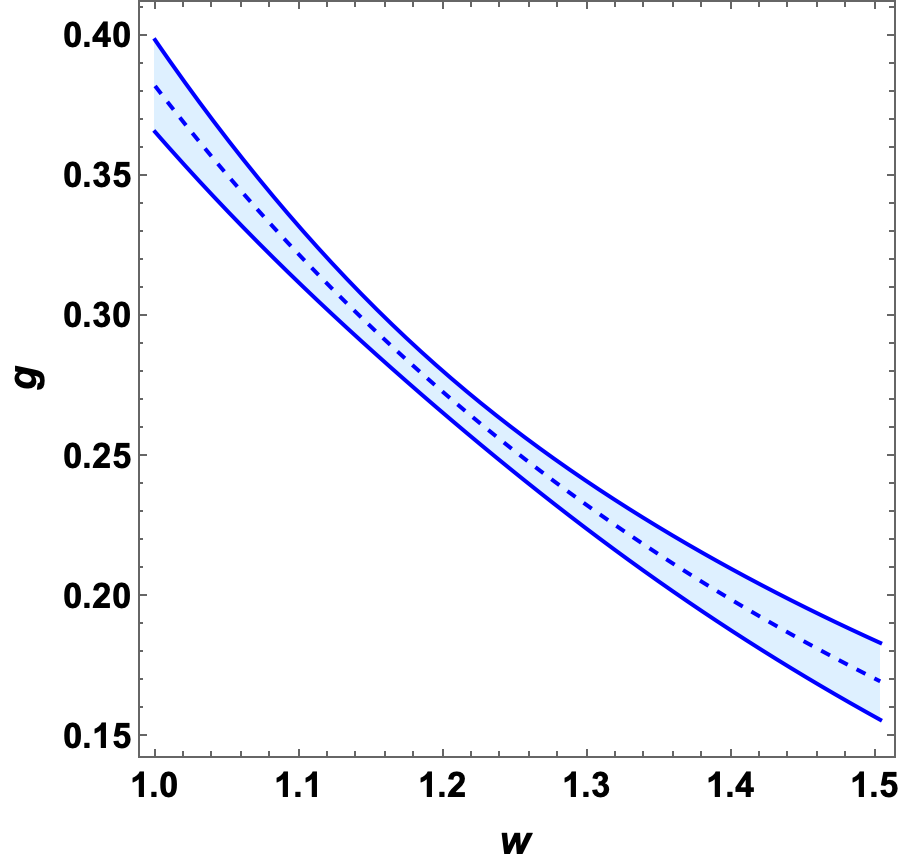} \\
 \includegraphics[scale=0.42]{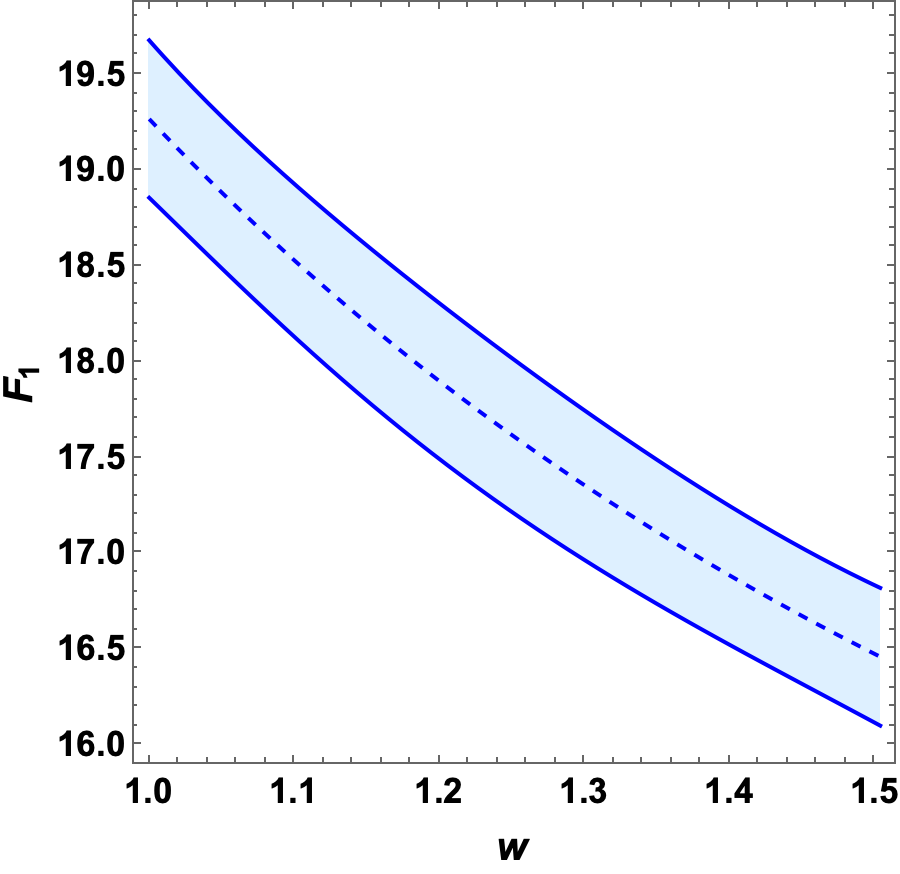} & \includegraphics[scale=0.43]{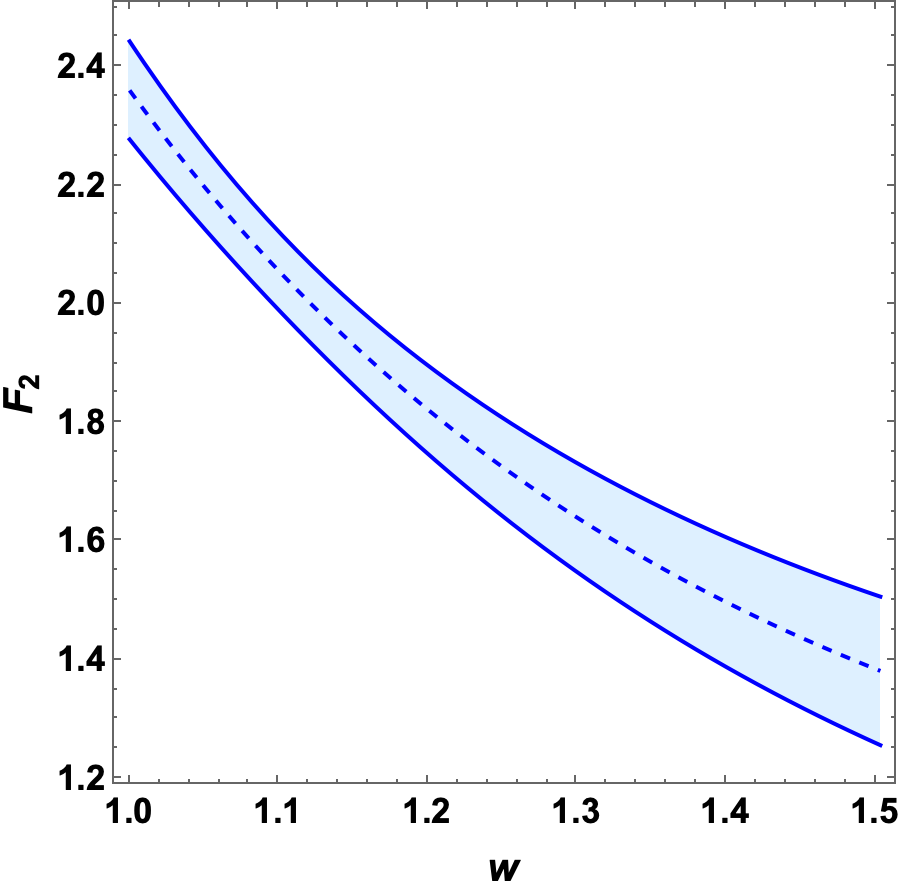} 
\end{tabular}
\end{center}
\caption{The $w$-dependence of the form factors using the fitted BGL parameters with the Belle data ~\cite{belle19} and the Fermilab-MILC lattice data~\cite{fermilab}. }
\label{fig:bellelatticeff_fm}
\end{figure}

{Next, we will perform the fit including the new lattice data given in Section~\ref{sec:lattice}. The new $\chi^2$ to minimise is the sum of Eq.(\ref{eq:chi2lattice}) and Eq.(\ref{eq:chi2belle}), i.e.
\begin{equation}
\chi^2_{\rm Belle+latt}(\vec{a}_{\rm BGL}, |V_{cb}|)=\chi^2_{\rm Belle} (\vec{a}_{\rm BGL}, |V_{cb}|)+\chi^2_{\rm latt}(\vec{a}_{\rm BGL})
\end{equation}
Now, we can determine both the form factor parameters $\vec{a}_{\rm BGL}$ and $|V_{cb}|$. 
It is important to mention that in this case, we also have to generate the MC sample for the lattice data, by fluctuating them \cn{according to their covariance matrix}, to perform the fit. The obtained fit results are given in Table~\ref{table:bellelatticeff}. Using the BGL parameters, we draw the $w$ dependence of the form factors, $f, g, F_1$ and $F_2$ (without prior) in Figs.~\ref{fig:bellelatticeff_jlqcd} and~\ref{fig:bellelatticeff_fm}.  We observe that the obtained bounds on $f$, and especially on $g$ form factors becomes very narrow in the whole region of $w$, as compared to Figs.~\ref{fig:latticeff} and ~\ref{fig:latticeffprior}. This comes from the fact that the Belle data constrains these parameters in the middle range of $w$ as seen in Fig.~\ref{fig:belleff}, while the lattice data constrain them in the lower $w$ region. As a result these two form factors become well constrained in the whole range of $w$.}

\begin{figure}[h!]
\begin{center}
\begin{tabular}{cc}
\multicolumn{1}{c}{\hspace{1.2cm} Belle+JLQCD} &  \multicolumn{1}{c}{\hspace{1cm} Belle+Fermilab-MILC} \\
 \includegraphics[scale=0.45]{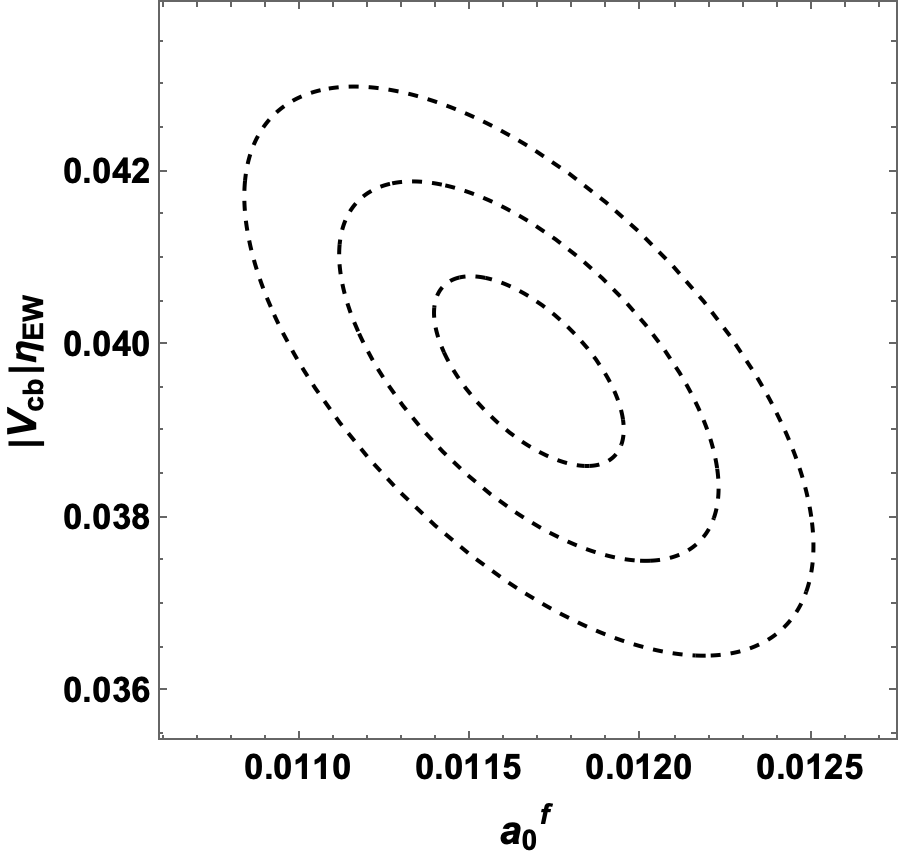} & 
 \includegraphics[scale=0.45]{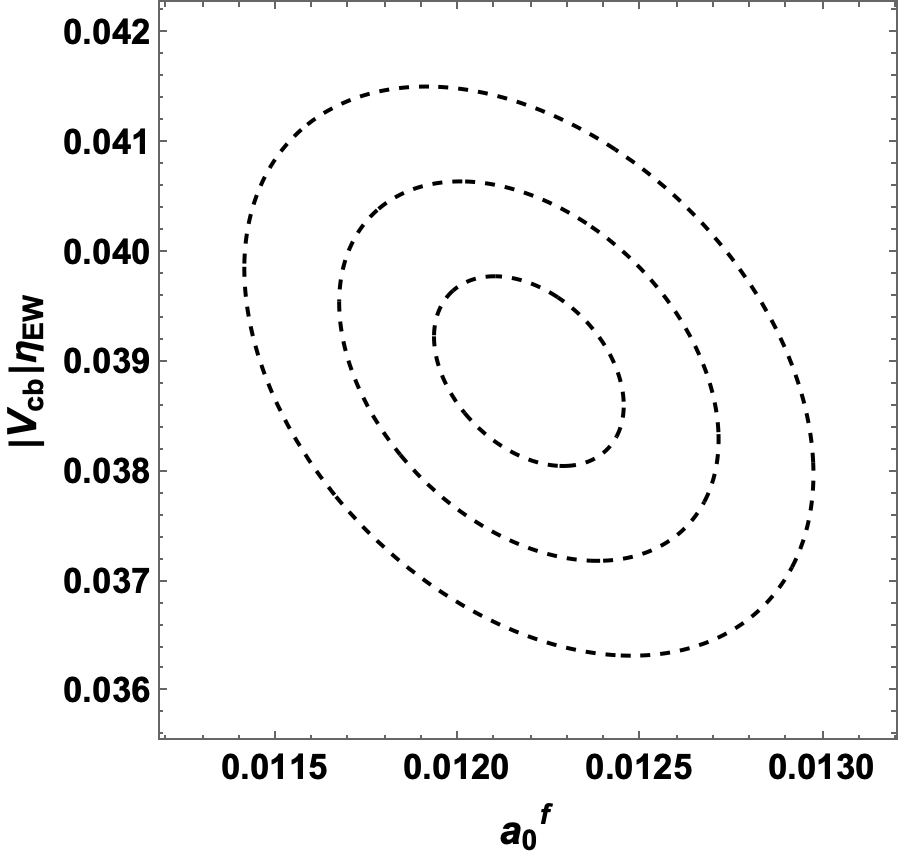}  
\end{tabular}
\end{center}
\caption{Correlation plots between $a^f_0$ and $|V_{cb}|\eta_{\rm EW}$, from combined fit of  Belle data~\cite{belle19} and lattice data. The first plot is with JLQCD data~\cite{jlqcd} and the second plot is with Fermilab-MILC data~\cite{fermilab}. The contours indicate $1\sigma$, $2\sigma$ and $3\sigma$ ranges.}
\label{fig:bellelatticecorr}
\end{figure}

{Concerning $|V_{cb}|$, with the $\eta_{\rm EW}=1.0066$, we find 
\begin{align}
\begin{aligned}
|V_{cb}|= 0.0394\pm 0.0011 \quad\quad\quad& {\rm JLQCD} \\
|V_{cb}|= 0.0386\pm 0.0009 \quad\quad\quad& {\rm Fermilab-MILC}
\end{aligned}
\end{align}
As expected, the $|V_{cb}|$ is strongly correlated to the value of $a^f_0$, as depicted in Fig.~\ref{fig:bellelatticecorr}. 
} 
\begin{table}
\begin{center}
\begin{tabular}{|c || d{13} | d{13} |}
\hline
\multicolumn{1}{|c||}{Fit parameters} & \multicolumn{1}{c|}{JLQCD} & \multicolumn{1}{c|}{Fermilab-MILC} \\
\hline\hline
$a^f_0 $ &0.01167(23)(16) & 0.01219(17)(19) \\
$a^f_1 $ & 0.0428(70)(69) &  0.0233(54)(105) \\
$a^f_2 $ & -0.47(20)(26) & -0.16(12)(17) \\
\hline
$a^g_0 $ & 0.02876(93)(64) & 0.03229(99)(96)  \\
$a^g_1 $ & -0.061(33)(25) & -0.166(35)(37) \\
$a^g_2$ & -0.0165(80)(87) & -0.03(24)(22) \\
\hline
$a^{F_1}_1 $ & 0.0089(21)(21) & 0.0030(13)(18) \\
$a^{F_1}_2 $ & -0.109(44)(37) & 0.002(26)(25) \\
\hline
$a^{F_2}_0 $ & 0.0515(18)(18) & 0.0519(12)(13) \\
$a^{F_2}_1 $ & -0.03(12)(13) & -0.175(55)(67) \\
\hline
$|V_{cb}|\eta_{\rm EW} $ & 0.03969(86)(67) & 0.03892(52)(69)  \\
\hline
\end{tabular}
\end{center}
\caption{The BGL form factors and $|V_{cb}|\eta_{\rm EW}$ obtained by a simultaneous fit of the Belle data~\cite{belle19} and the lattice  data by JLQCD~\cite{jlqcd} (second column) and Fermilab-MILC~\cite{fermilab} (third column) collaborations. The first error is statistical and the second is systematic.}
\label{table:bellelatticeff}
\end{table}

\subsection{Test of new physics by fitting the Belle and lattice data: example of right-handed model }\label{sec:binned_npfit}
{In this section, we suggest the simultaneous fit of Belle and lattice data to test the NP model with right-handed current as an example. Strictly speaking, we cannot use the currently available Belle data to perform this fit since its correlation assumes the SM as mentioned earlier. For this reason, our fit result is not completely reliable, though it may give a useful insight in the future analysis for Belle II. }
\par
\begin{figure}[h!]
\begin{center}
\begin{tabular}{cc}
 \includegraphics[scale=0.46]{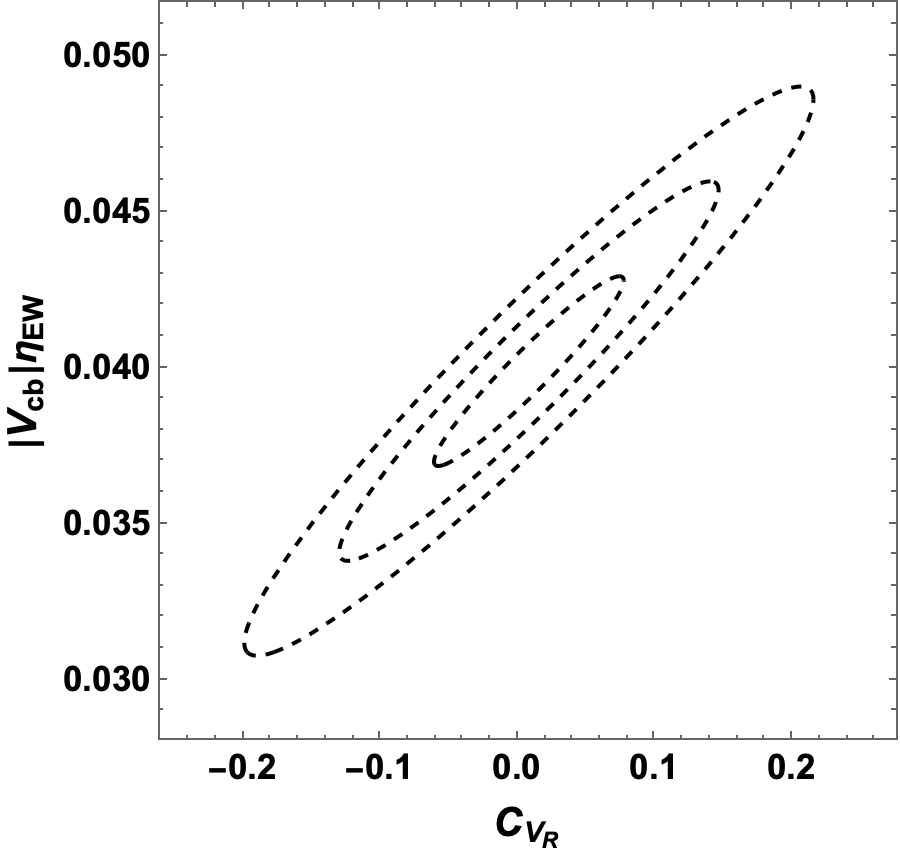} & 
 \includegraphics[scale=0.45]{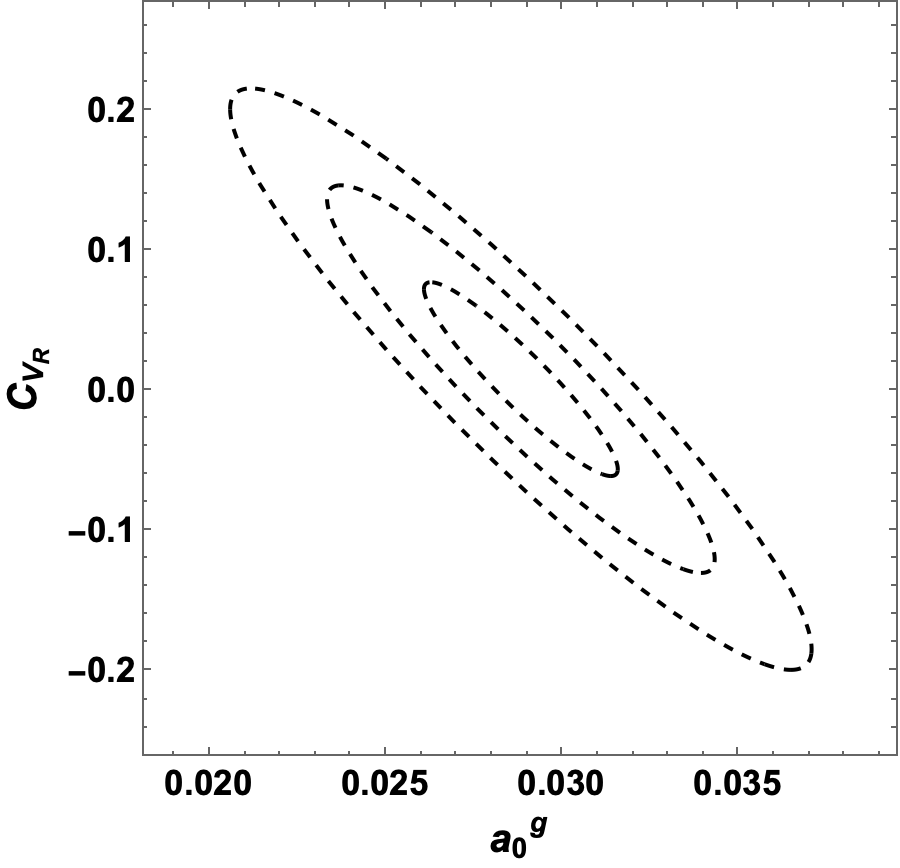} \\
\multicolumn{2}{c}{ \includegraphics[scale=0.45]{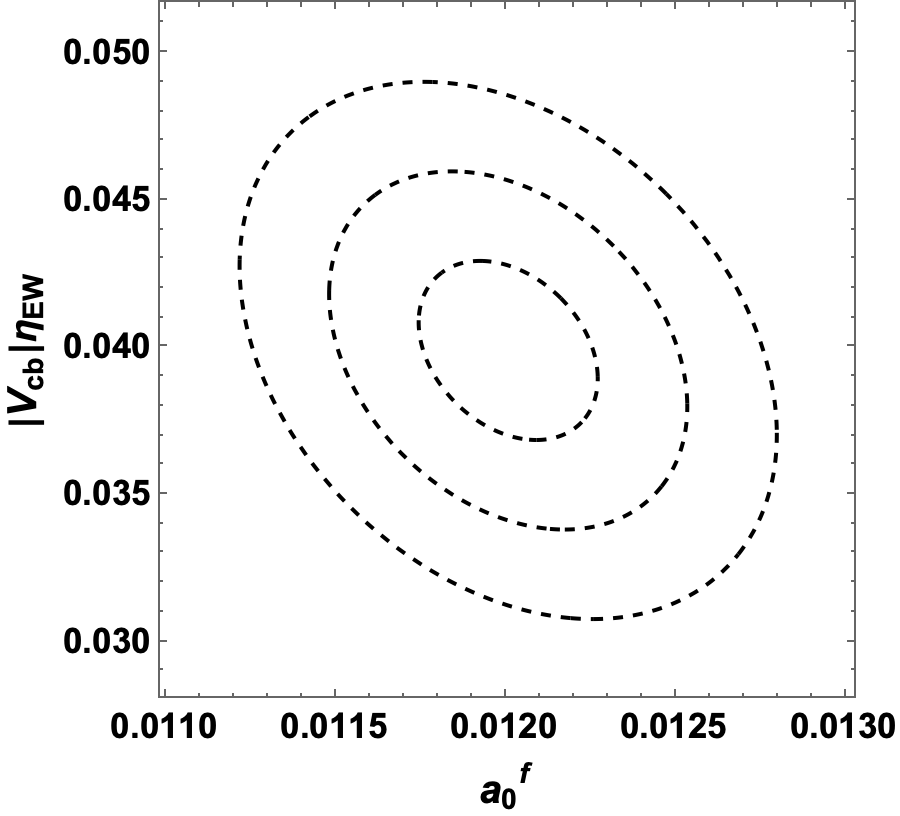} }  
\end{tabular}
\end{center}
\caption{Correlation plots between $a^f_0$, $|V_{cb}|\eta_{\rm EW}$ and $C_{\rm V_R}$, from combined fit of  Belle data~\cite{belle19} and JLQCD lattice data~\cite{jlqcd}. $C_{ V_R}$ is assumed to be real. The contours indicate $1\sigma$, $2\sigma$ and $3\sigma$ ranges.}
\label{fig:bellelatticerhcorr_jlqcd}
\end{figure}
\begin{figure}[h!]
\begin{center}
\begin{tabular}{cc}
\includegraphics[scale=0.45]{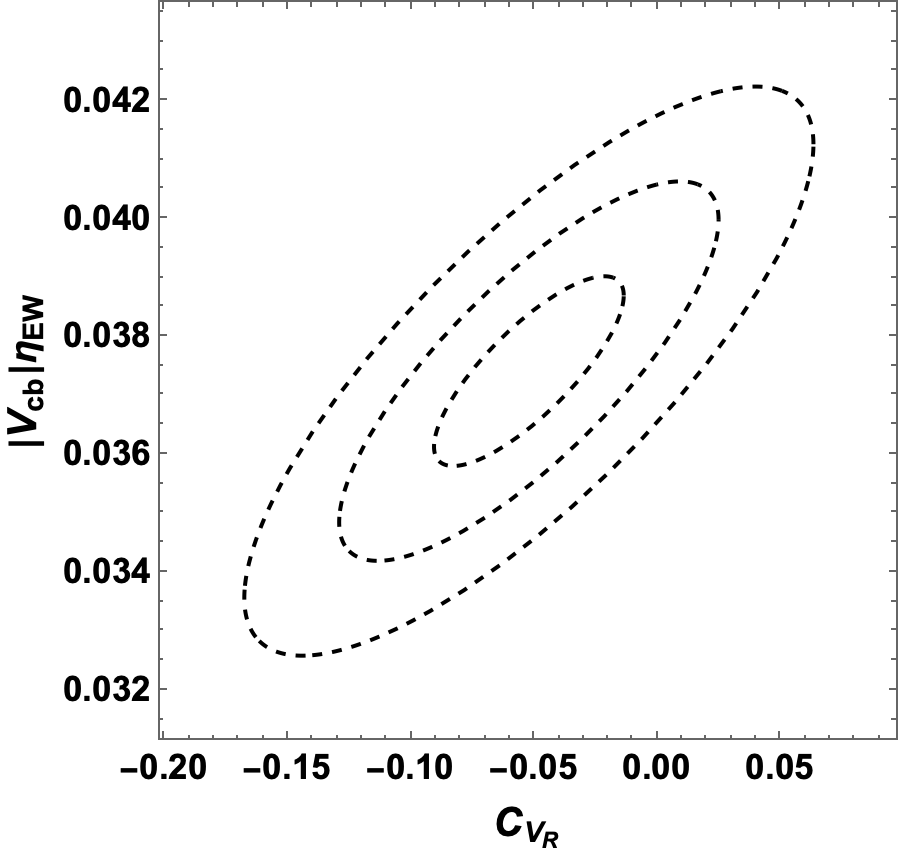} & \includegraphics[scale=0.47]{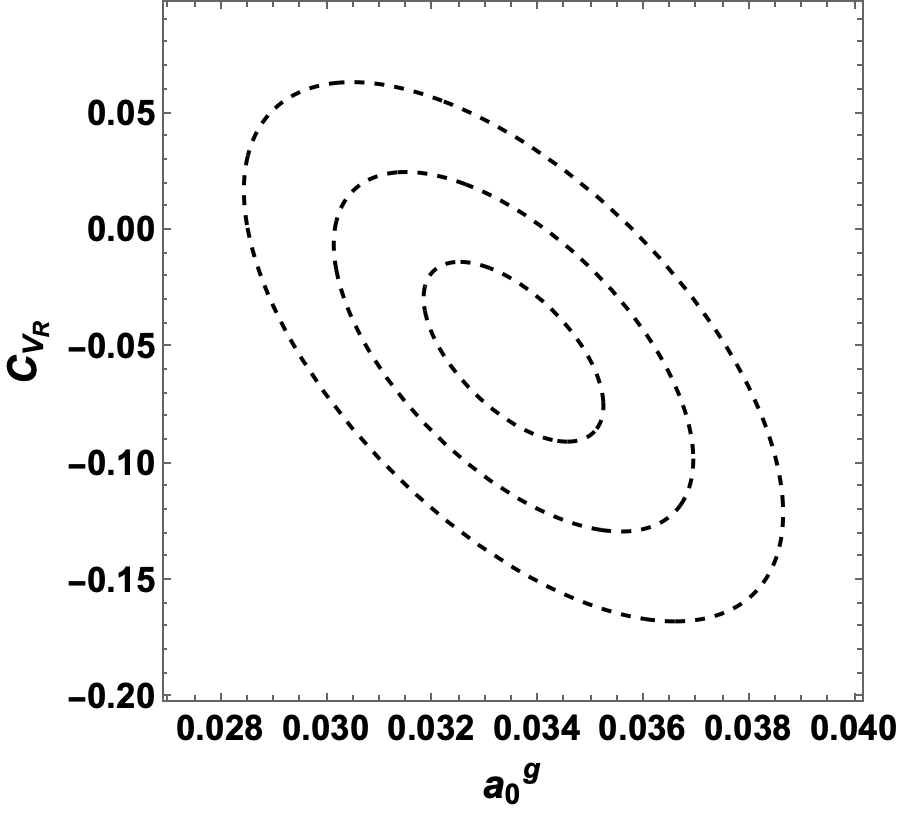} \\
\multicolumn{2}{c}{ \includegraphics[scale=0.45]{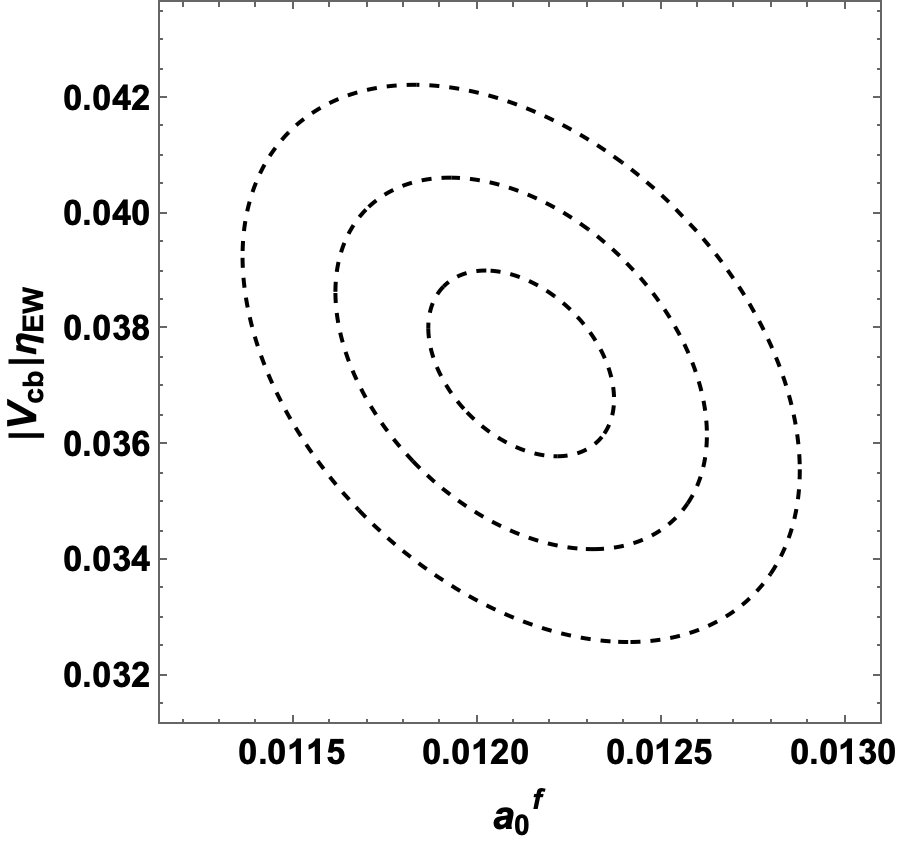} }
\end{tabular}
\end{center}
\caption{Correlation plots between $a^f_0$, $a^g_0$,  $|V_{cb}|\eta_{\rm EW}$ and $C_{\rm V_R}$, from combined fit of  Belle data~\cite{belle19} and Fermilab-MILC lattice data~\cite{fermilab}. $C_{\rm V_R}$ is assumed to be real. The contours indicate $1\sigma$, $2\sigma$ and $3\sigma$ ranges.}
\label{fig:bellelatticerhcorr_fm}
\end{figure}

With the right-handed contribution, the $J_i$ functions which describe the differential decay rate in Eq.~\eqref{eq:rateJ} include a new parameter $C_{\rm V_R}$ as given in Table~\ref{table:jfunclhrh}. In general $C_{\rm V_R}$ can be a complex number, but we assume it to be a real parameter for simplicity in this section. Now the $\chi^2$ \cn{used in the} fit is given as
\begin{align}
\chi^2_{\rm Belle+latt}(\vec{a}_{\rm BGL}, |V_{cb}|, C_{\rm V_R})=\chi^2_{\rm Belle} (\vec{a}_{\rm BGL}, |V_{cb}|, C_{\rm V_R})+\chi^2_{\rm latt}(\vec{a}_{\rm BGL})
\end{align}
where $C_{\rm V_R}$ is included in the $N_i^{\rm exp}$ in  $\chi^2_{\rm Belle}$. 
The obtained fit results are given in Table~\ref{table:bellelatticerhff}. 
\cn{The three largest correlations are between $a_0^f-|V_{cb}|$, $a_0^g-C_{\rm V_R}$ and $|V_{cb}|-C_{\rm V_R}$, which are shown in Figs.~\ref{fig:bellelatticerhcorr_jlqcd} and~\ref{fig:bellelatticerhcorr_fm}.} Indeed, the central value of $a_0^g$ for two lattice inputs disagree as seen in, for example, Table~\ref{table:latticeff}, and that seems to be the reason for slightly different $C_{\rm V_R}$ obtained in the fit.  As the correlation between  $|V_{cb}|$ and $C_{\rm V_R}$ is positive, the negative $C_{\rm V_R}$ obtained for Fermilab-MILC lowers the value of $|V_{cb}|$. As mentioned in the introduction, the SM $|V_{cb}|$ is expected to be higher than the nominal fit, this contradicts to the global picture.

\begin{table}
\begin{center}
\begin{tabular}{| c || d{13} | d{13}| }
\hline
\multicolumn{1}{|c||}{Fit parameters} & \multicolumn{1}{c|}{JLQCD} & \multicolumn{1}{c|}{Fermilab-MILC} \\
\hline\hline
$a^f_0 $ & 0.01200(20)(18)  & 0.01212(19)(17) \\
$a^f_ 1$ & 0.0193(34)(69) & 0.0193(29)(74) \\
$a^f_ 2$ &  -0.036(69)(51) & -0.066(25)(26) \\
\hline
$a^g_0 $ &  0.0288(19)(20) & 0.0335(12)(12) \\
$a^g_1 $ & -0.070(34)(40)  & -0.123 (46)(49) \\
$a^g_ 2$ & -0.0051(71)(59)   & -0.0060(18)(20)  \\
\hline
$a^{F_1}_ 1$ & 0.0051(13)(19)  & 0.0026(11)(15) \\
$a^{F_1}_2 $ & -0.049(27)(27)  & 0.004(23)(22) \\
\hline
$a^{F_2}_ 0$ & 0.0481(18)(18)  & 0.0522(12)(12) \\
$a^{F_2}_ 1$ & 0.030(60)(67)  & -0.193(50)(52)  \\
\hline
$C_{\rm V_R} $ &  0.008(47)(51) & -0.052(27)(28)  \\
$ |V_{cb}|\eta_{\rm EW}$ & 0.0399(21)(22) & 0.0374(11)(12) \\
\hline
\end{tabular}
\end{center}
\caption{The BGL form factors, $C_{\rm V_R}$ and $|V_{cb}|\eta_{\rm EW}$ obtained by a simultaneous fit of the Belle data~\cite{belle19} and the lattice  data by JLQCD~\cite{jlqcd} (second column) and Fermilab-MILC~\cite{fermilab} (third column) collaborations. The first error is statistical and the second is systematic.}
\label{table:bellelatticerhff}
\end{table}

\section{Unbinned angular analysis of \BDlnu }\label{sec:unbinned}
{In this section, following the method introduced in~\cite{emizh}, we  investigate the sensitivity of the NP in the unbinned method. The Belle analysis is limited to the binned analysis as of today, while once higher statistics will be available, a possibility of the unbinned analysis will open. The unbinned analysis, with full kinematical information, unlike the binned method which uses only integrated one dimension information, should provide a more detailed information.  In this section, \cn{we investigate how the unbinned method may provide additional information with respect to the binned method studied in the previous section.} }

{As unbinned data is not available from Belle, we synthesise the data first: we use all the fitted parameters by Belle experiment in~\cite{belle19} (form factors and $V_{cb}$) as truth values, and reproduce the differential decay rate of \BDlnu decay. Then, to compensate for the limited statistics, we first separate the data in $w$ bins of 10 equal intervals. Then, for each $w$ bin, we use the 3D unbinned distribution ($\cos\theta_v, \cos\theta_\ell, \chi$). This is the same strategy applied for the $P_i^{(\prime)}$ measurements at LHCb in~\cite{lhcbp5}. }

{Comparing to our study in~\cite{emizh}, we have three new inputs for this work; i) we use new lattice data introduced in Section~\ref{sec:lattice}, ii) the NP model includes not only right-handed current but also pseudoscalar and tensor terms, iii) we include the branching ratio to the fit.
The latter point is important in order to determine the NP parameters (i.e. $C_{\rm V_R, P, T}$) and $V_{cb}$ simultaneously. In our previous work, we attempted to use the  number of signal of \BDlnu process for this purpose, but to compare its theoretical predictions to the experimental value, it would require the knowledge experimental efficiency. Using the branching ratio provides a much simpler solution. 
} 

\subsection{Unbinned maximum likelihood method}
{
 Let us now illustrate how to perform the unbinned fit with maximum likelihood method.} We start by obtaining the {normalised Probability Density Function (PDF)} in terms of the 11 independent angular observables $J_i$. For this, {we first give the expression of decay rate integrated over all the angles}
\begin{equation}
\frac{d\Gamma}{dw} = \frac{3M_B M_{D^*}^2}{4(4\pi)^2} G_F^2 |V_{cb}|^2 |\eta_{\rm EW}|^2\mathcal{B}(D^* \rightarrow D \pi) 
 \frac{8\pi}{9} \left( 6J'_{1s} + 3J'_{1c} - 2J'_{2s} - J'_{2c} \right)
\end{equation}
where $J'_i \equiv J_i \sqrt{w^2-1}(1 - 2wr + r^2)$. {Then, we integrate over each $w$-bin and define the integrated decay rate as }
\begin{equation}\label{decayratewbin}
\begin{aligned}
\langle \Gamma \rangle_{w-\rm bin} &= \frac{3M_B M_{D^*}^2}{4(4\pi)^2} G_F^2 |V_{cb}|^2 |\eta_{\rm EW}|^2 \mathcal{B}(D^* \rightarrow D \pi) \\
&\times \frac{8\pi}{9} \left( 6 \langle J'_{1s} \rangle_{ w-\rm bin} + 3  \langle J'_{1c} \rangle_{w-\rm bin} - 2 \langle J'_{2s} \rangle_{ w-\rm bin} - \langle J'_{2c} \rangle_{w-\rm bin} \right)
 \end{aligned}
\end{equation}
{where 
\[\langle J_i \rangle_{w-{\rm bin}}\equiv \int_{w-{\rm bin}} J_i dw\]}
Hereafter, the index "$w$-bin" is implicit. The normalised PDF {for each $w$-bin} now can be {written} by the new normalised angular coefficients {$\vec{\langle g\rangle}=\langle g_i\rangle$} as
\begin{align}\label{eq:normpdf}
\begin{aligned}
\hat{f}_{\langle \vec{g} \rangle} (\cos\theta_V, \cos\theta_\ell,\chi) &= \frac{9}{8\pi} 
 \Big\{ \frac{1}{6}(1 - 3 \langle g_{1c} \rangle + 2 \langle g_{2s} \rangle +\langle g_{2c} \rangle) \sin^2\theta_V + \langle g_{1c} \rangle \cos^2 \theta_V  \\
 &+ (\langle g_{2s} \rangle \sin^2 \theta_V  + \langle g_{2c} \rangle \cos^2 \theta_V )\cos 2\theta_l  \\
 &+ \langle g_{3} \rangle \sin^2 \theta_V \sin^2 \theta_l  \cos 2\chi_l  
 + \langle g_{4} \rangle \sin 2\theta_V \sin 2\theta_l \cos \chi_l  \\
 &+ \langle g_{5} \rangle \sin 2\theta_V \sin \theta_l  \cos \chi_l  
 + (\langle g_{6s} \rangle \sin^2 \theta_V  + \langle g_{6c} \rangle \cos^2 \theta_V ) \cos \theta_l  \\
 &+ \langle g_{7} \rangle \sin 2\theta_V  \sin \theta_l  \sin \chi_l  
 + \langle g_8 \rangle \sin 2\theta_V \sin 2\theta_l \sin \chi_l  \\
 &+ \langle g_{9} \rangle \sin^2 \theta_V  \sin^2 \theta_l  \sin 2\chi_l  \Big\}
\end{aligned}
\end{align}
where
\begin{equation} \label{eq:gifunc}
\langle g_i \rangle \equiv \frac{\langle J'_i \rangle}{6 \langle J'_{1s} \rangle + 3 \langle J'_{1c} \rangle - 2\langle J'_{2s} \rangle - \langle J'_{2c} \rangle}
\end{equation}
{Then, we can write the log-likelihood to determine the angular coefficients $\langle \vec{g}\rangle $ from the data: }
\begin{equation}\label{eq:likelihood}
\mathcal{L}(\langle \vec{g} \rangle) = \sum_{i=1}^{N} \ln \hat{f}_{\langle \vec{g} \rangle}(e_i)
\end{equation}
where $e_i$ denotes the experimental events in ($\cos\theta_V, \cos\theta_\ell,\chi$) and $N$ are the number of events for each $w$-bin. 
The best fit value for $\langle g_i \rangle$ is obtained by maximising the likelihood 
\begin{equation}
\left. \frac{\partial \mathcal{L}}{\partial \langle g_i \rangle}\right|_{\langle g_i \rangle=\langle g_i \rangle_{\rm fit}} =0
\end{equation}
The inverse of the covariance matrix is obtained as
\begin{equation}\label{covmatinverse}
V^{-1}_{ij}=-\left.\frac{\partial^2 \mathcal{L}}{\partial \langle g_i \rangle\partial \langle g_j \rangle}\right|_{\langle g_{i} \rangle=\langle g_{i} \rangle_{\rm fit}}
\end{equation}
\cn{Eq.~\eqref{covmatinverse} provides the 11x11 matrix for each $w$-bin. There are 10 such matrices since we have 10 bins in $w$.}

\subsection{Fitting the theory parameters with the pseudo-data together with the lattice inputs}

{In this work, instead of performing a Toy Monte Carlo with generated events, we obtain directly $\langle g_i \rangle_{\rm fit}$ as well as $V_{ij}$ from the normalised PDF  $\hat{f}_{\langle \vec{g} \rangle}$ with {\it truth} inputs (i.e. the fitted form factors and $V_{cb}$ in~\cite{belle19} as mentioned earlier). The detailed procedure is given in the Appendix~\ref{app:statprocedure}. }

{Now, we can fit the theory parameters to the 'experimentally' obtained angular coefficients and their correlations  by using the following $\chi^2$:
\begin{align}\label{eq:chi2angle}
&\chi^2_{\rm angle}(\vec{a}_{\rm BGL}, C_{\rm NP}) = \\
&\sum_{w-{\rm bin} = 1}^{10} \left[ N_{w-{\rm bin}}\hat{V}^{-1}_{ij} \left( \langle g_i^{\rm exp}  \rangle - \langle g_i^{\rm th} (\vec{a}_{\rm BGL}, C_{\rm NP}) \rangle\right) \left( \langle g_j^{\rm exp}  \rangle - \langle g_j^{\rm th} (\vec{a}_{\rm BGL}, C_{\rm NP}) \rangle\right) \right]_{w-{\rm bin}} \nonumber
\end{align}
}{where $\hat V$ is the scaled covariance matrix, $N_{w-{\rm bin}}$ are the number of events in the $w$-bin} {and $\langle g_i^{\rm exp} \rangle$ is the best fit value of the angular coefficients  (see Appendix~\ref{app:statprocedure}  for details) while $\langle g_i^{\rm th} (\vec{a}_{\rm BGL}, C_{\rm NP})\rangle$ are the theoretical expressions, which depend on the form factors and NP Wilson coefficients. We assume about the same amount of signal events as in~\cite{belle19}. Note that as $\chi^2_{\rm angle}$ term is made from the normalised angular coefficients, the overall constants such as $V_{cb}$ get cancelled.}

{In order to constrain from the total number of events, we use the world average of the branching ratio, as mentioned earlier: 
\[\mathcal{B}^{\rm exp}(B\to D^*\ell\nu)=0.0495\pm 0.0011\]
Then, the $\chi^2$ from the branching ratio measurement can be written as, 
\begin{equation}
\chi^2_{\mathcal{B}}(\vec{a}_{\rm BGL},V_{cb}, C_{\rm NP}) = \left(\frac{\mathcal{B}^{\rm th}(\vec{a}_{\rm BGL}, V_{cb}, C_{\rm NP})-0.0495}{0.0011}\right)^2
\end{equation}
where $\mathcal{B}^{\rm th}(\vec{a}_{\rm BGL}, V_{cb}, C_{\rm NP})$ is the theoretical prediction of the branching ratio including NP contribution.}

{Finally adding the lattice constraint, the total $\chi^2$ to fit is
\begin{align}
\chi^2_{\rm unbinned}(\vec{a}_{\rm BGL}, V_{cb}, C_{\rm NP})=\chi^2_{\rm angle}(\vec{a}_{\rm BGL}, C_{\rm NP})+\chi^2_{\rm latt}(\vec{a}_{\rm BGL})+\chi^2_{\mathcal{B}}(\vec{a}_{\rm BGL}, V_{cb}, C_{\rm NP})
\end{align}
In the following, we show the fit results. We assume single non-zero NP coupling at a time (i.e. $C_{\rm V_R}\neq 0, C_{\rm P}\neq 0$ or $C_{\rm T}\neq 0$) and we also assume that they are real number. }

\subsubsection{Right-handed model: $C_{\rm V_R}\neq 0$}
\begin{figure}
\begin{center}
\begin{tabular}{cc}
\includegraphics[scale=0.5]{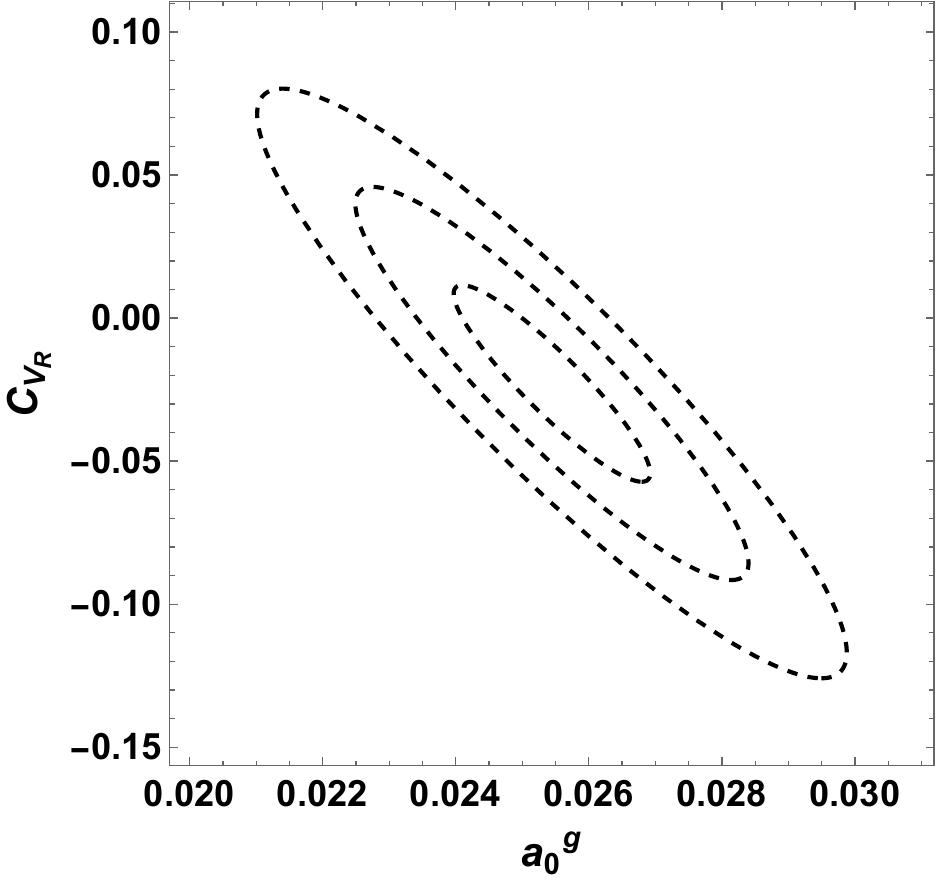} & \includegraphics[scale=0.5]{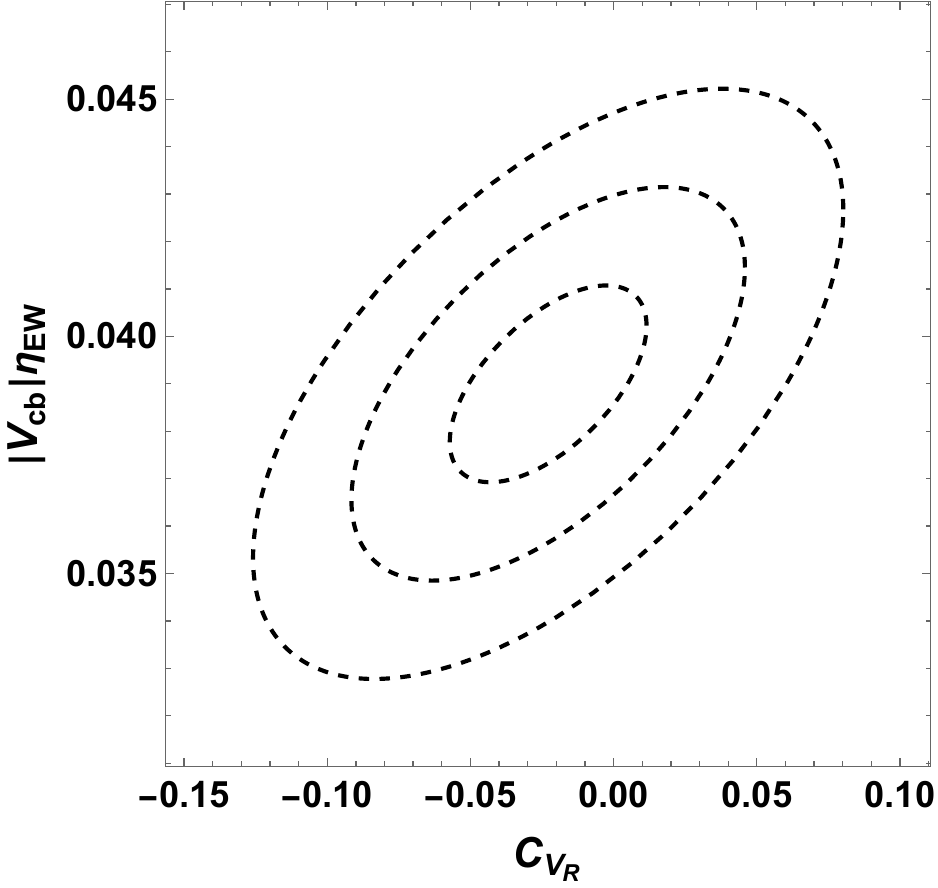} 
\end{tabular}
\end{center}
\caption{Correlation plots between $a^g_0$, $|V_{cb}|\eta_{\rm EW}$ and $C_{\rm V_R}$, from combined fit of the pseudo-data and JLQCD lattice data~\cite{jlqcd}. $C_{\rm V_R}$ is assumed to be real. The contours indicate $1\sigma$, $2\sigma$ and $3\sigma$ ranges.}
\label{fig:unbinnedrhcorr_jlqcd}
\end{figure}

\begin{figure}
\begin{center}
\begin{tabular}{cc}
\includegraphics[scale=0.5]{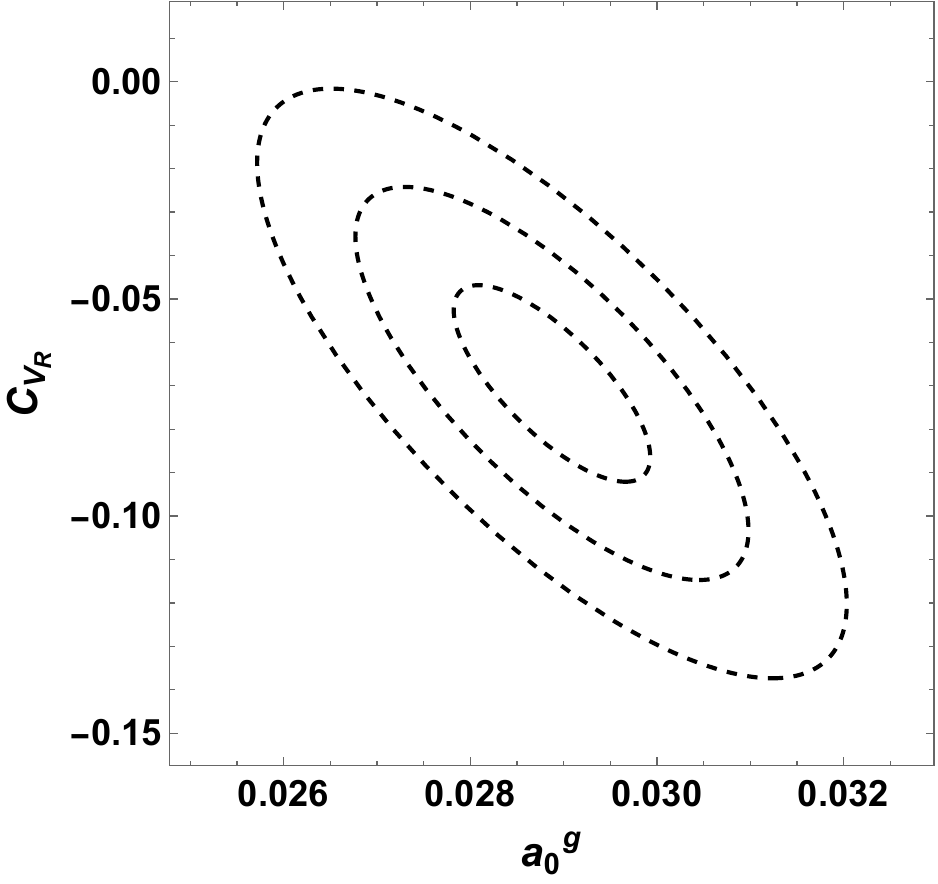} & \includegraphics[scale=0.5]{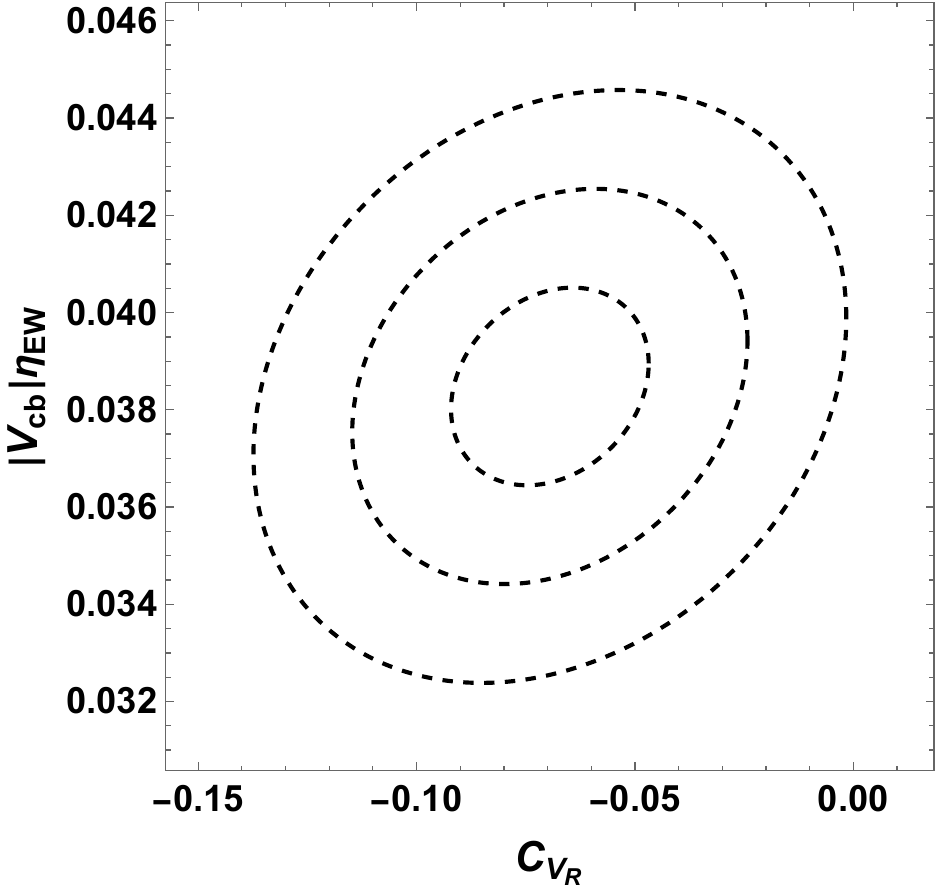} 
\end{tabular}
\end{center}
\caption{Correlation plots between $a^g_0$, $|V_{cb}|\eta_{\rm EW}$ and $C_{\rm V_R}$, from combined fit of the pseudo-data and Fermilab-MILC lattice data~\cite{fermilab} $C_{\rm V_R}$ is assumed to be real. The contours indicate $1\sigma$, $2\sigma$ and $3\sigma$ ranges.}
\label{fig:unbinnedrhcorr_fm}
\end{figure}

{The fitted result is given in Table~\ref{table:unbinnedrh}. 
Let us first compare this unbinned analysis result with the binned one given in Table~\ref{table:bellelatticerhff}. As our pseudodata does not have systematic error, we can compare only the statistical one (i.e. the first parenthesis in Table ~\ref{table:bellelatticerhff}). As expected, the statistical errors are \cn{smaller} for the unbinned analysis though, hardly a factor of 2 for $C_{\rm V_R}$. Figs.~\ref{fig:unbinnedrhcorr_jlqcd} and~\ref{fig:unbinnedrhcorr_fm} show the correlations, which are also very similar to the binned study in Figs.~\ref{fig:bellelatticerhcorr_jlqcd} and Fig.~\ref{fig:bellelatticerhcorr_fm}. The \cn{Wilson coefficient} $C_{\rm V_R}$ is more significantly different from zero in this case.  }
\begin{table}
\begin{center}
\begin{tabular}{| c || d{9} | d{9}| }
\hline
\multicolumn{1}{|c||}{Fit parameters} & \multicolumn{1}{c|}{JLQCD} & \multicolumn{1}{c|}{Fermilab-MILC} \\
\hline\hline
$a^f_0$ & 0.01294(20) & 0.01327(20) \\
$a^f_1$ &  0.018(11)& -0.005(13) \\
$a^f_2$ &  -0.21(53)& -0.28(61) \\
\hline
$a^g_0 $ & 0.0255(15) & 0.0289(11) \\
$a^g_1 $ & -0.045(35)& -0.130(49) \\
$a^g_2 $ & 0.4(11)& 1.0(14) \\
\hline
$a^{\mathcal{F}_1}_1$ & 0.0061(22) & -0.0001(23) \\
$a^{\mathcal{F}_1}_2 $ & -0.095(94) & -0.05(10) \\
\hline
$a^{\mathcal{F}_2}_0$ & 0.0483(14) & 0.0537(13) \\
$a^{\mathcal{F}_2}_1$ &-0.004(50)  & -0.276(62) \\
\hline
$ C_{\rm V_R}$ & -0.023(34) & -0.070(23) \\
$|V_{cb}| \eta_{\rm EW}$ & 0.0390(21) & 0.0385(20) \\
\hline
\end{tabular}
\end{center}
\caption{The BGL form factors, $C_{\rm V_R}$ and $|V_{cb}|\eta_{\rm EW}$ obtained by a simultaneous fit of the pseudo-data and the lattice  data by JLQCD~\cite{jlqcd} (second column) and Fermilab-MILC~\cite{fermilab} (third column) collaborations. The errors are statistical only.}
\label{table:unbinnedrh}
\end{table}

\subsubsection{Pseudoscalar model: $C_{\rm P}\neq 0$}
\begin{figure}
\begin{center}
\begin{tabular}{cc}
 \includegraphics[scale=0.5]{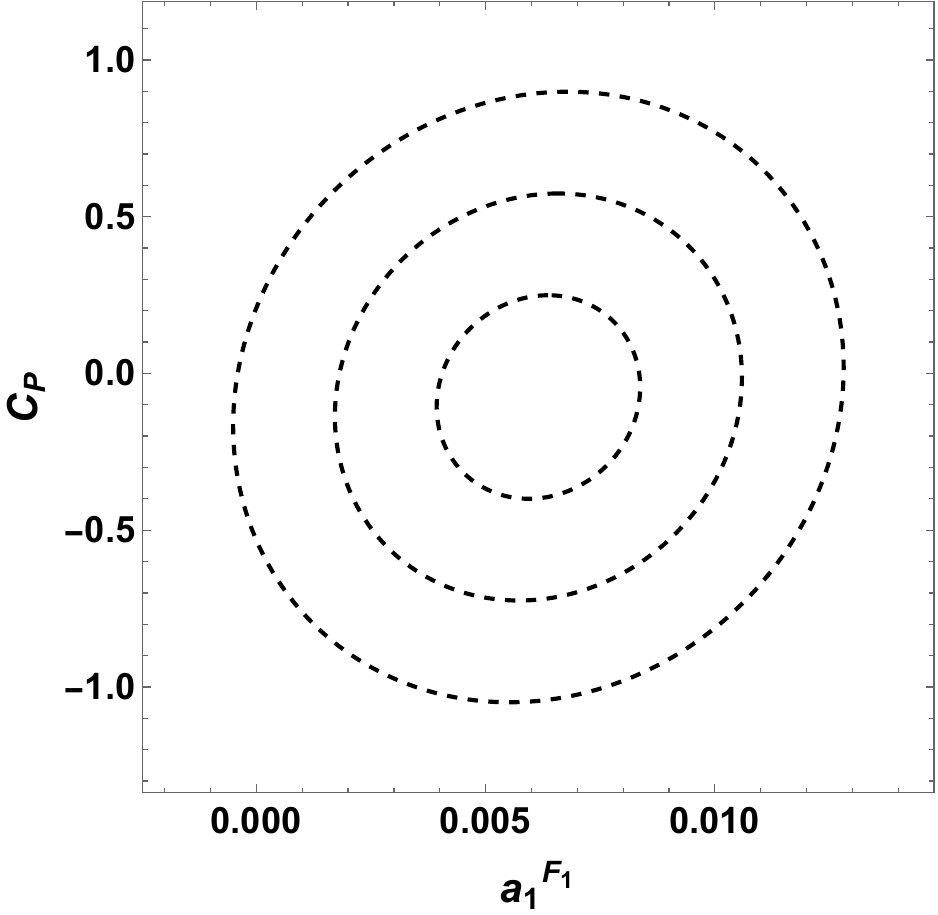} & \includegraphics[scale=0.52]{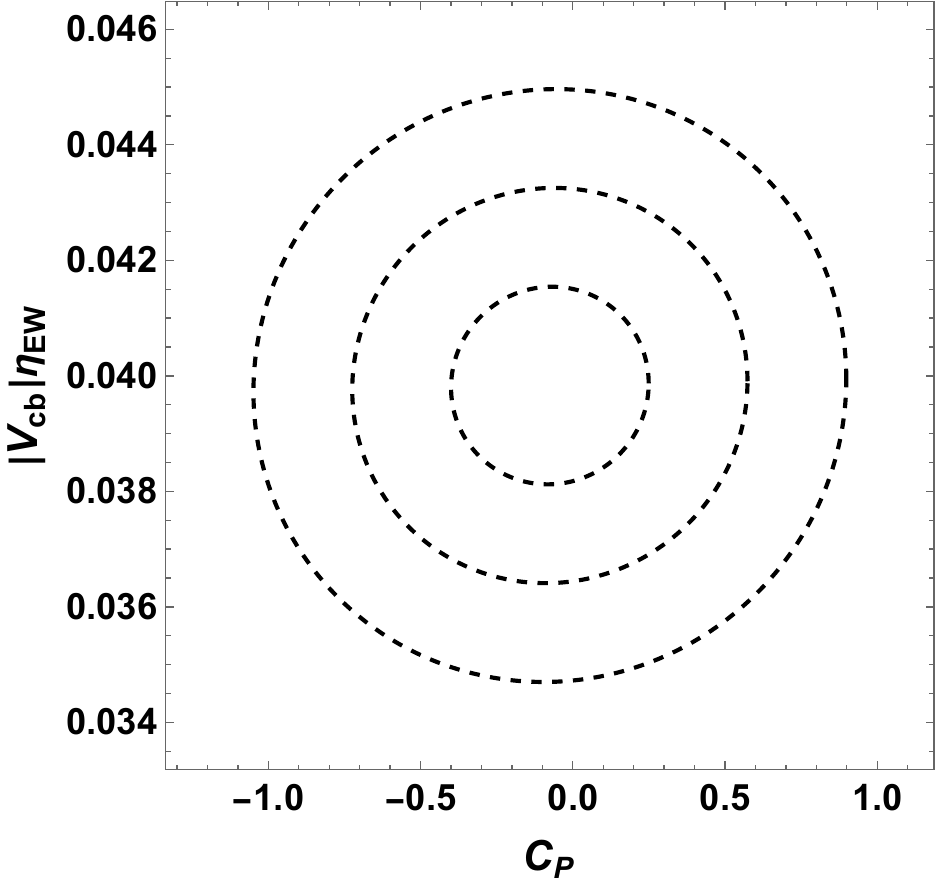} 
\end{tabular}
\end{center}
\caption{Correlation plots between $a^{\mathcal{F}_1}_1$, $|V_{cb}|\eta_{\rm EW}$ and $|C_{\rm P}|$, from combined fit of the pseudo-data and JLQCD lattice data~\cite{jlqcd}. The contours indicate $1\sigma$, $2\sigma$ and $3\sigma$ ranges.}
\label{fig:unbinnedpscorr_jlqcd}
\end{figure}

\begin{figure}
\begin{center}
\begin{tabular}{cc}
\includegraphics[scale=0.5]{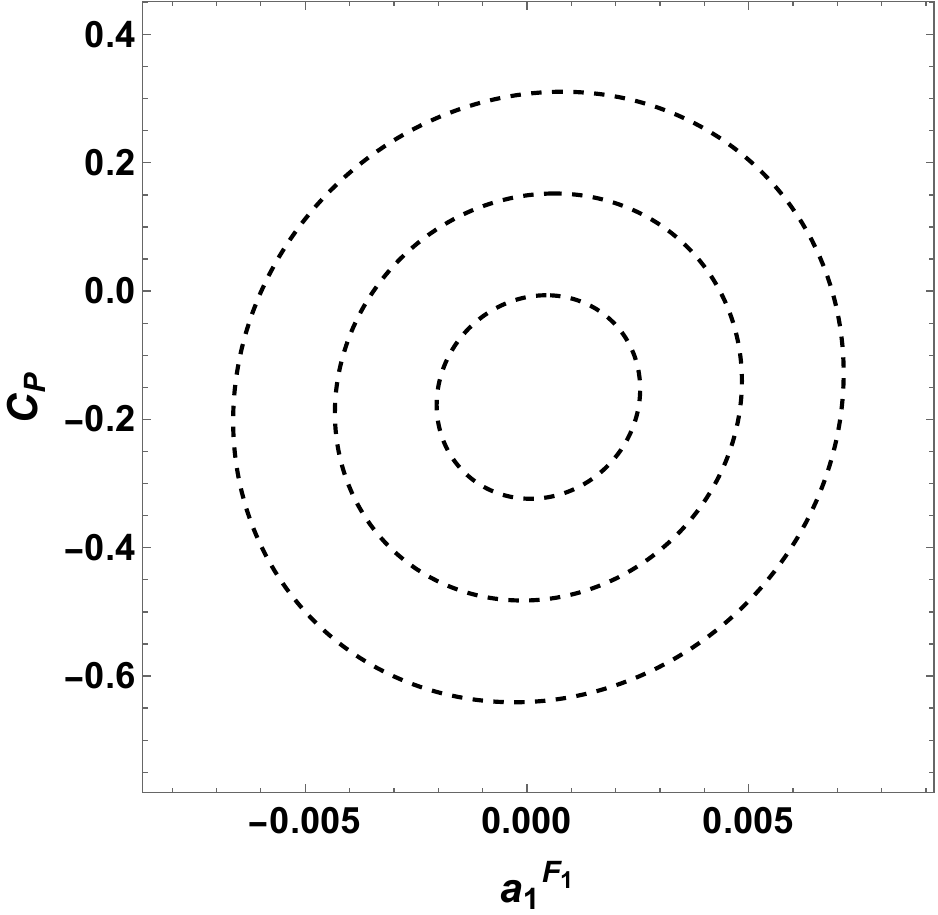}  & \includegraphics[scale=0.5]{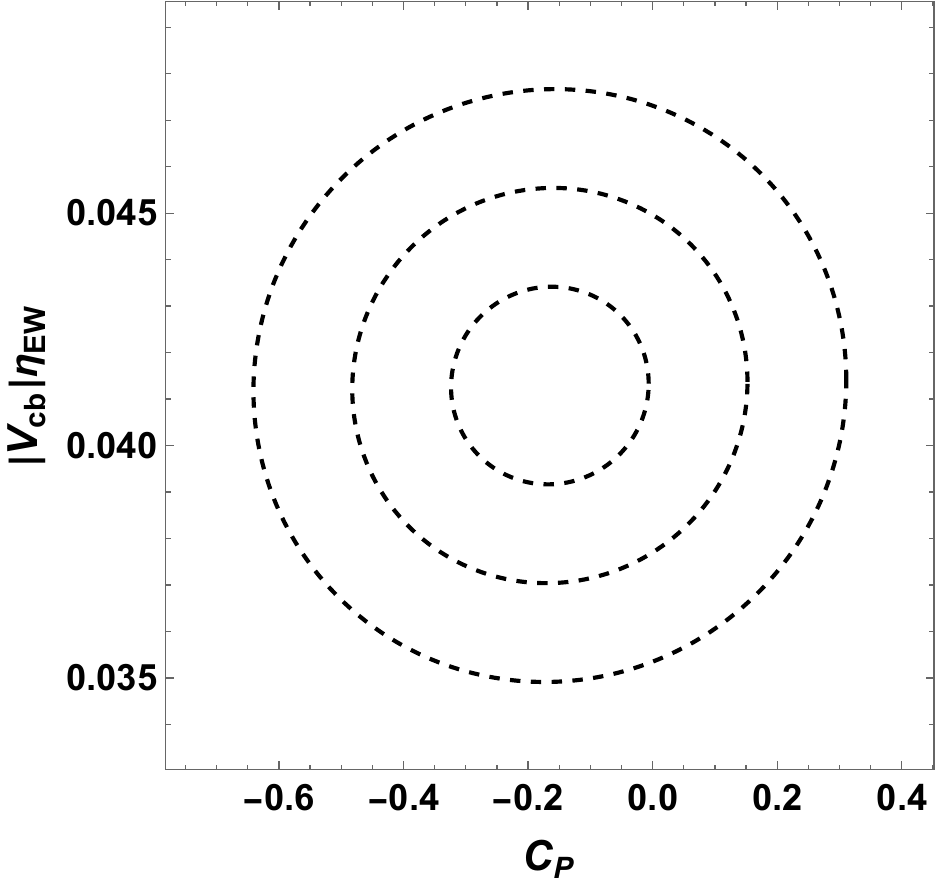} 
\end{tabular}
\end{center}
\caption{Correlation plots between $a^{\mathcal{F}_1}_1$, $|V_{cb}|\eta_{\rm EW}$ and $|C_{\rm P}|$, from combined fit of the pseudo-data and Fermilab-MILC lattice data~\cite{fermilab}. The contours indicate $1\sigma$, $2\sigma$ and $3\sigma$ ranges.}
\label{fig:unbinnedpscorr_fm}
\end{figure}

{As we can see in Tables~\ref{table:jfunclhrh} and~\ref{table:jfuncpt}, the $C_{\rm P}$ parameter enters only to the angular observable $J_{1c}$ in squared form, $|C_{\rm P}|^2$. There is no interference with the SM term (i.e. $C_{\rm V_L}$,) as such terms are proportional to the lepton mass, which is neglected.  As a result the sensitivity to this parameter is quite poor. The fitted result is given in Table~\ref{table:unbinnedpseudoscalar} where we find that $|C_{\rm P}|$ is constrained only at 20-30\% precision. 
As mentioned earlier, the pseudoscalar form factor is obtained by relating it to the axial vector form factor. This could potentially induce \cn{a bias} from the uncertainties of charm and quark masses. However, we find that this error is negligible here: we tested this by inflating the errors in the quark masses by $\sim 1 \rm GeV$. 
The results on $|C_{\rm P}|$ with JLQCD and Fermilab-MILC inputs are slightly different. As shown in Figs.~\ref{fig:unbinnedpscorr_jlqcd} and~\ref{fig:unbinnedpscorr_fm}, there is a small correlation between $C_{\rm P}$ and $a_1^{\mathcal{F}_1}$ whose values are different for these two inputs and this might be the reason of this difference. Also shown in Figs.~\ref{fig:unbinnedpscorr_jlqcd} and~\ref{fig:unbinnedpscorr_fm}, the correlation between $|C_{\rm P}|$ and $|V_{cb}|$ is very small.  Thus, a shift in $|C_{\rm P}|$ would not influence the value of $V_{cb}$ as seen in the the case of $C_{\rm V_R}\neq 0$. } 

\begin{table}
\begin{center}
\begin{tabular}{|c || d{9}| d{9}| }
\hline
\multicolumn{1}{|c||}{Fit parameters} & \multicolumn{1}{c|}{JLQCD} & \multicolumn{1}{c|}{Fermilab-MILC} \\
\hline\hline
$a^f_0$ & 0.01294(20) & 0.01336(20) \\
$a^f_1$ &  0.018(11)& -0.003(13) \\
$a^f_2$ &  -0.20(53)& -0.47(61) \\
\hline
$a^g_0 $ & 0.02456(60) & 0.02626(70) \\
$a^g_1 $ & -0.057(29)& -0.154(45) \\
$a^g_2 $ & 0.6(11)& 0.9(12) \\
\hline
$a^{\mathcal{F}_1}_1$ & 0.0062(22) & -0.0003(23) \\
$a^{\mathcal{F}_1}_2 $ & -0.095(94) & -0.08(10) \\
\hline
$a^{\mathcal{F}_2}_0$ & 0.0484(14) & 0.0530(13) \\
$a^{\mathcal{F}_2}_1$ &-0.002(50)  & -0.271(61) \\
\hline
$  |C_{\rm P}|$ & -0.08(33) & -0.17(16) \\
$|V_{cb}|\eta_{\rm EW}$ & 0.0398(17) & 0.0413(21) \\
\hline
\end{tabular}
\end{center}
\caption{The BGL form factors, $|C_{\rm P}|$ and $|V_{cb}|\eta_{\rm EW}$ obtained by a simultaneous fit of the pseudo-data and the lattice  data by JLQCD~\cite{jlqcd} (second column) and Fermilab-MILC~\cite{fermilab} (third column) collaborations. The errors are statistical only.}
\label{table:unbinnedpseudoscalar}
\end{table}

\subsubsection{Tensor model: $C_{\rm T}\neq 0$}
\begin{figure}
\begin{center}
\begin{tabular}{cc}
\includegraphics[scale=0.52]{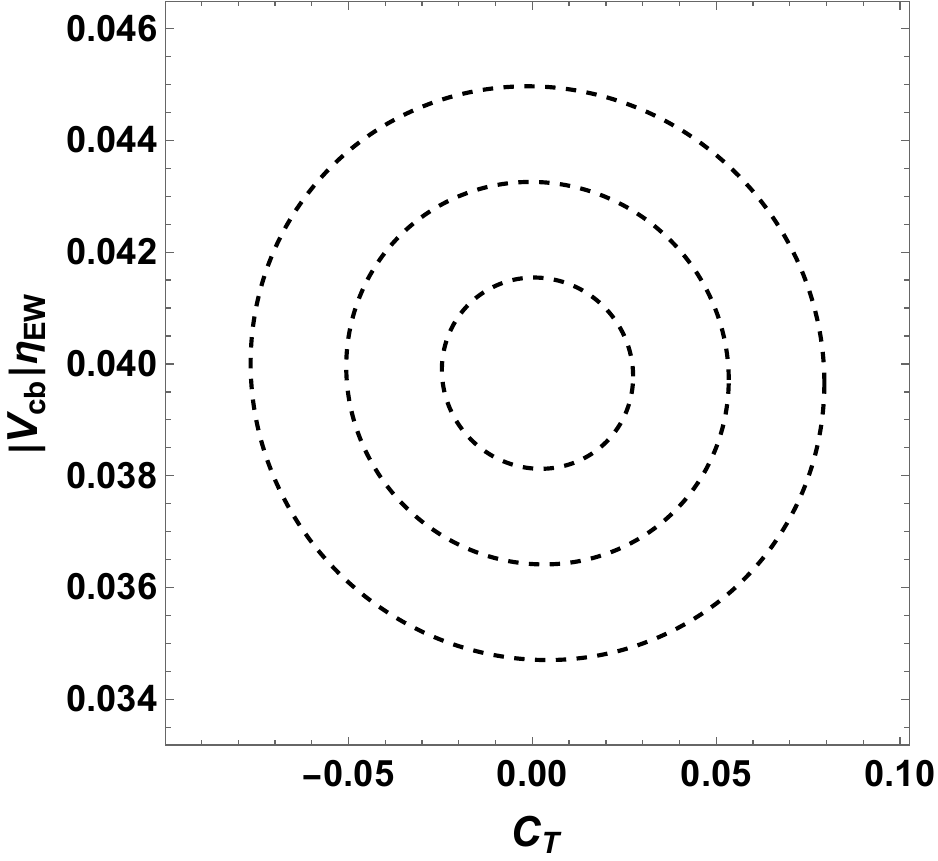} &
\includegraphics[scale=0.5]{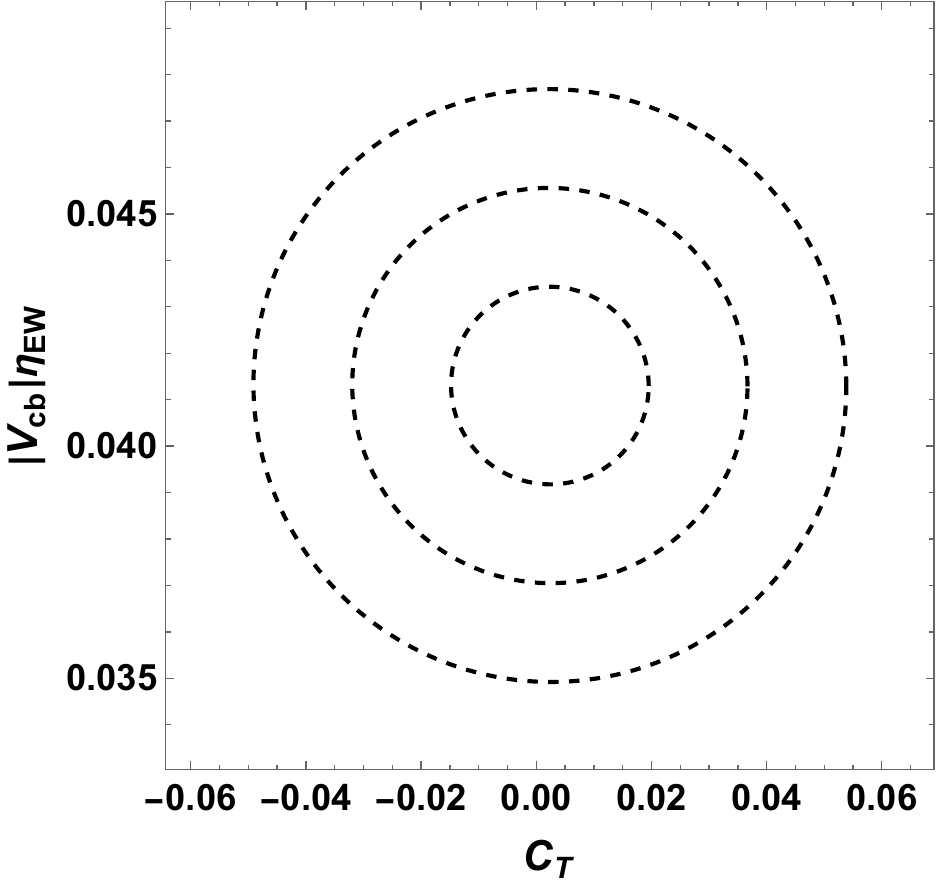} 
\end{tabular}
\end{center}
\caption{Correlation plots between  $|V_{cb}|\eta_{\rm EW}$ and $|C_{\rm T}|$, from combined fit of the pseudo-data and JLQCD lattice data (left)~\cite{jlqcd} and Fermilab-MILC lattice data (right)~\cite{fermilab}. The contours indicate $1\sigma$, $2\sigma$ and $3\sigma$ ranges.}
\label{fig:unbinnedtcorr_tensor}
\end{figure}

{In the case of $C_{\rm T}\neq 0$ model, several $J_i$ functions depend on $C_{\rm T}$, though, similarly to the $C_{\rm P}\neq 0$ model, the interference to the SM term cancels due to the small lepton masses (see Tables~\ref{table:jfunclhrh} and~\ref{table:jfuncpt}).  
The fitted results are given in Table~\ref{table:unbinnedtensor} where we find that $|C_{\rm T}|$ is much more constrained than $|C_{\rm P}|$, at 2-3\% level. The result does not change after taking into account the uncertainties from charm and quark masses, which appear in the formula relating the tensor and vector form factors.  In this case, we observe almost no correlation between $|C_{\rm T}|$ and form factors and the correlation between $|C_{\rm T}|$ and $V_{cb}$ is also small as well (see  Figs.~\ref{fig:unbinnedtcorr_tensor}). } 

\begin{table}
\begin{center}
\begin{tabular}{|c ||d{9} |d{9}| }
\hline
\multicolumn{1}{|c||}{Fit parameters} & \multicolumn{1}{c|}{JLQCD} & \multicolumn{1}{c|}{Fermilab-MILC} \\
\hline\hline
$a^f_0$ & 0.01294(20) & 0.01336(20) \\
$a^f_1$ &  0.018(11)& -0.003(13) \\
$a^f_2$ &  -0.21(54)& -0.48(61) \\
\hline
$a^g_0 $ & 0.02456(60) & 0.02626(70) \\
$a^g_1 $ & -0.057(29)& -0.154(45) \\
$a^g_2 $ & 0.6(11)& 0.9(12) \\
\hline
$a^{\mathcal{F}_1}_1$ & -0.0062(22) & -0.0003(23) \\
$a^{\mathcal{F}_1}_2 $ & -0.095(94) & -0.08(10) \\
\hline
$a^{\mathcal{F}_2}_0$ & 0.0484(14) & 0.0530(13) \\
$a^{\mathcal{F}_2}_1$ &-0.002(50)  & -0.271(61) \\
\hline
$ |C_{\rm T}|$ & 0.001(26) & 0.002(17) \\
$|V_{cb}|\eta_{\rm EW}$ & 0.0398(17) & 0.0413(21) \\
\hline
\end{tabular}
\end{center}
\caption{The BGL form factors, $|C_{\rm T}|$ and $|V_{cb}|\eta_{\rm EW}$ obtained by a simultaneous fit of the pseudo-data and the lattice  data by JLQCD~\cite{jlqcd} (second column) and Fermilab-MILC~\cite{fermilab} (third column) collaborations. The errors are statistical only.}
\label{table:unbinnedtensor}
\end{table}

\section{Conclusions}\label{sec:conclusions}
{In this article, we investigated the potential of the angular analysis of  \BDlnu {decay} to search for NP in the light of the new lattice QCD data on the form factors.  Previously, such an analysis was not possible, as the form factors, the SM CKM parameter $V_{cb}$ and the NP parameters could not be fitted simultaneously from the experimental data.  Starting with an effective Hamiltonian containing different NP operators (left/right-handed, pseudoscalar and tensor), we first computed the complete angular distribution of \BDlnu decay, which confirms the previous computation~\cite{london}. While the lattice data provide only the vector and axial-vector form factors, the  pseudoscalar and tensor form factors can be derived from the equation of motion, up to the uncertainties in the charm and bottom quark masses. We use the lattice data from two collaborations, Fermilab-MILC and JLQCD,  which provide  the four form factors at three different $w$ points in the low recoil region. We first extrapolated the form factors to the higher $w$ region based on the BGL parametrisation and showed that the errors on them at high $w$ region are large even after taking into account the unitary bound. }

{Next, we used the Belle data \cite{belle19} and fit then together with the lattice data. The result shows that  the errors on the form factors in both low and high $w$ region reduces significantly. The SM parameter, $V_{cb}$, appears as an overall factor and it can not be fitted from the experimental data only. We have fitted $V_{cb}$ with the new lattice data and found $|V_{cb}|=(39.4 \pm 1.1)\times 10^{-3}$ (JLQCD) and $|V_{cb}|=(38.6 \pm 0.9)\times 10^{-3}$ (Fermilab-MILC), which agrees with  previous studies. In order to evaluate whether  Belle data (one dimensional and binned) can further constrain  NP parameters, we next introduced the right-handed current, $C_{\rm V_R}$. Strictly speaking, this fit is not possible as of today as  Belle data assume SM correlations. Nevertheless, we performed the fit with $C_{\rm V_R}\neq 0$ for demonstration. We found a surprisingly high sensitivity of this binned Belle data to $C_{\rm V_R}$, at the level of 4-7\%. We also found that $C_{\rm V_R}$ is strongly correlated to the vector form factor parameter $a_0^g$ and the current slight discrepancy between JLQCD and Fermilab-MILC on this parameter leads to different conclusions: the latter shows some hint of a negative shift of $C_{\rm V_R}$. On the other hand, as $C_{\rm V_R}$ is positively correlated  to the $|V_{cb}|$ value,  this lowers  $V_{cb}$, which is not favourable from the global fit point of view. In any case, our result demonstrated that the Belle (II) binned fit can provide valuable information on NP and we suggest Belle (II) collaboration to perform such fits. }
 
{ In the last section, we have investigated the possibility of the unbinned fit. Using synthesised data based on the same Belle dataset together with  lattice inputs, we have performed a sensitivity study on the right-handed, pseudoscalar and tensor types of NP models. For the right-handed case, we find very similar results as the binned analysis, with the sensitivity slightly higher. The pseudoscalar and tensor models are expected to be less constrained as their interference with SM is proportional to the final state lepton mass and it is suppressed in the $\ell=e, \mu $ cases. Nevertheless, we find that the pseudoscalar and tensor couplings, $|C_{\rm P,T}|$, are constrained, respectively, at 20-30\% and 2-3\% levels. For these cases, different from the right-handed case, the correlation between the NP couplings and $|V_{cb}|$ is very small. Thus, a  nonzero observation of these couplings would not conflict with the global fit result of $|V_{cb}|$. }

\section*{Acknowledgements}
{We would like to acknowledge T. Kaneko, Ph. Urquijo, D. Ferlewicz and E. Waheed for their collaboration at the the early stage of this project. 
We thank F. Le Diberder for his  advice and careful reading of the manuscript. We also thank R. Mizuk for his helpful comments.  This work was supported by TYL-FJPPL. \cn{Z.R.Huang was supported in part by the National Natural Science Foundation of China under Grant 12305104.}}

\appendix
\section{Relationship between pseudoscalar/tensor and vector form factors}\label{app:ffrelations}
We use equations of motion in order to express pseudoscalar and tensor form factors in terms of vector form factors \cite{tensorff}. 
\\\\
\underline{Pseudoscalar form factors}:
\\

Using the equation of motion
\begin{equation}
i\partial_{\mu} (\bar{c}\gamma^{\mu} \gamma^5 b) =- (m_b + m_c)\bar{c}\gamma^5 b
\end{equation}
the pseudo scalar matrix event becomes
\begin{equation}
\begin{aligned}
\langle D^* (p_{D^*},\epsilon_2) | \bar{c}\gamma^5 b| \bar{B}(p_B) \rangle &= \frac{-1}{m_b + m_c} q_{\mu} \langle D^* (p_{D^*},\epsilon_2) | \bar{c}\gamma^{\mu} \gamma^5 b| \bar{B}(p_B) \rangle \\
\end{aligned}
\end{equation}
where $m_{b,c}$ are $b$- and $c$- quark masses, respectively, and $q^{\mu} = p_B^{\mu}-p_{D^*}^{\mu}$. In our numerical analysis we use $m_b = 4.18^{+0.03}_{-0.02}$ GeV and $m_c = 1.27 \pm 0.02$ GeV \cite{pdg22}.
\\\\
\underline{Tensor form factors}
\\

The guiding equation of motion is given by
\begin{equation}\label{eq:tensoreom}
\partial_{\mu}(\bar{c}\sigma^{\mu\nu}b) = -(m_b + m_c)\bar{c}\gamma^{\nu}b
\end{equation}
which gives us the equations
\begin{equation}\label{eq:tensoreommatrix}
-i q_{\mu} \langle D^* (p_{D^*},\epsilon_2) | \bar{c}\sigma^{\mu\nu} b| \bar{B}(p_B) \rangle = -(m_b + m_c) \langle D^* (p_{D^*},\epsilon_2) |\bar{c}\gamma^{\nu}b| \bar{B}(p_B) \rangle 
\end{equation}
Note that there exists two scalar current terms in Eq.~\eqref{eq:tensoreom}, which we do not write, as they get cancelled when we solve for the form factors. Eq.~\eqref{eq:tensoreommatrix} equation gives the same relation for $+$ and $-$ polarization of $D^*$, and both sides go to zero for longitudinal (0) polarization. Therefore, we have only two equations in total, and we need two more to determine all the tensor form factors (including the form factor for $B \rightarrow D$ decay). For this, we have the pseudo tensor equation of motion given by
\begin{equation}
\partial_{\mu}(\bar{c}\sigma^{\mu\nu}\gamma^5 b) = (m_b - m_c)\bar{c}\gamma^{\nu} \gamma^5 b
\end{equation}
where the pseudo tensor matrix elements can be related to the tensor matrix elements by the relation $\bar{c}\sigma_{\mu\nu} \gamma^5 b = -\frac{i}{2} \epsilon_{\mu\nu\alpha\beta} \bar{c}\sigma^{\alpha\beta}b$ with the convention $\epsilon_{0123} = 1$. This relation gives us the following equation
\begin{equation}
-\frac{q_{\mu}}{2} \epsilon^{\mu\nu\alpha\beta} \langle D^* (p_{D^*},\epsilon_2) | \bar{c}\sigma_{\alpha\beta} b| \bar{B}(p_B) \rangle = (m_b - m_c) \langle D^* (p_{D^*},\epsilon_2) | \bar{c}\gamma^{\nu} \gamma^5 b| \bar{B}(p_B) \rangle
\end{equation}
where we get two equations - one from $\epsilon_2(\pm)$ and other from $\epsilon_2(0)$. Now these four equations can be solved to write the expressions of the four tensor form factors ($h_{T, T_1, T_2, T_3}$) in terms of vector form factors  ($h_{+, -, V, A_1, A_2, A_3}$), which are given as follows \cite{tensorff}:

\begin{align}\label{eq:hqettensorvectorrelation}
\begin{aligned}
h_T(w)& = \frac{m_b + m_c}{m_B + m_D} \Big[ h_+(w) - \frac{1+r_D}{1-r_D}h_-(w) \Big] \\
h_{T_1}(w)& = \frac{1}{2(1+r_{D^*}^2 - 2 r_{D^*}w)} \Big[ \frac{m_b - m_c}{m_B - m_{D^*}} (1-r_{D^*})^2(w+1) h_{A_1}(w)   \\
& - \frac{m_b + m_c}{m_B + m_{D^*}} (1+r_{D^*})^2(w-1) h_{V}(w)  \Big] \\
h_{T_2}(w)&= \frac{(1-r_{D^*}^2)(w+1)}{2(1+r_{D^*}^2 - 2 r_{D^*}w)} \Big[  \frac{m_b - m_c}{m_B - m_{D^*}} h_{A_1}(w) -  \frac{m_b + m_c}{m_B + m_{D^*}}  h_{V}(w)  \Big] \\
h_{T_3}(w) &= \frac{-1}{2(1+r_{D^*})(1+r_{D^*}^2 - 2 r_{D^*}w)} \Big[ 2  \frac{m_b - m_c}{m_B - m_{D^*}} r_{D^*}(w+1)h_{A_1}(w) \\
&+  \frac{m_b - m_c}{m_B - m_{D^*}} (1+r_{D^*}^2 - 2 r_{D^*}w) (h_{A_3}(w) - r_{D^*}h_{A_2}(w)) \\
&- \frac{m_b + m_c}{m_B + m_{D^*}} (1+r_{D^*})^2 h_{V}(w) \Big]
\end{aligned}
\end{align}
In addition, the BGL form factors are related to the HQET form factors as follows:
\begin{equation}\label{eq:bglhqet}
\begin{aligned}
g &= \frac{h_V}{M_B \sqrt{r}} \\
f &= M_B \sqrt{r} (1+w) h_{A_1} \\
\mathcal{F}_1 &= M_B^2 \sqrt{r} (1+w) \left[ (w-r)h_{A_1} - (w-1)(r h_{A_2}+ h_{A_3}) \right] \\
\mathcal{F}_2 &= \frac{1}{\sqrt{r}} \left[ (1+w)h_{A_1} +(rw -1)h_{A_2} + (r-w)h_{A_3}\right]
\end{aligned}
\end{equation}
We can solve Eq.~\eqref{eq:bglhqet} and Eq.~\eqref{eq:hqettensorvectorrelation} to write the tensor form factors in terms of BGL parameters, which gives us the tensor helicity amplitudes given in Eq.~\eqref{helampbgl}.

\section{Statistical procedure}\label{app:statprocedure}
This method of generating the covariance matrices from the truth values is based on the fact that the statistical error scales as $\sqrt{N_{\rm event}}$, where $N_{\rm event}$ are the number of events.
\par 
Starting with the normalied PDF $\hat{f}_{\langle \vec{g} \rangle}$ given in Eq.~\eqref{eq:normpdf}, the covariance matrix for $\langle \vec{g} \rangle$ functions is given by
\begin{align}
V_{ij}^{-1} = N_{\rm event} \int\  \left.\left(\frac{\partial \hat{f}(x)_{\langle \vec{g}_i \rangle}}{\partial \langle \vec{g}_i \rangle}  \frac{\partial \hat{f}(x)_{\langle \vec{g}_j \rangle}}{\partial \langle \vec{g}_j \rangle}  \frac{1}{\hat{f}(x)_{\langle \vec{g}_i \rangle}}\right)\right|_{\langle \vec{g}_i \rangle=\langle \vec{g}_i \rangle^{\rm exp}}\   dx 
\end{align}
where 
\begin{itemize}
\item $\langle \vec{g}_i \rangle$ is the vector of observables.
\item $\langle \vec{g}_i \rangle ^{\rm exp}$ is the vector of truth values, which we obtain from the experimentally measured form factors in \cite{belle19}.
\item $dx$ represents integration over the complete phase space, i.e. the three angles that describe the decay. 
\item $N_{\rm event}$ is the number of events in the corresponding $w$-bin: we have 10 $w$-bins, and thus, we have 10 such covariance matrices , the dimension of each being $11\times11$ (corresponding to 11 independent $\langle {g}_i \rangle$ functions).
\end{itemize}
{We can define a rescaled covariance matrix $\hat{V}$ as 
\begin{align}
\hat{V} =N_{\rm event} V 
\end{align}
}
Using this pseudo dataset, we can perform a maximum likelihood estimate with our model assumptions (right-handed, pseudoscalar, tensor etc.), using the $\chi^2$ given in Eq.~\eqref{eq:chi2angle}.

\bibliographystyle{JHEP}
\bibliography{bibliographybldnpaper.bib}

\end{document}